\documentclass[12pt]{iopart}
\usepackage{iopams}
\usepackage{amsmath}
\usepackage{graphicx}
\usepackage{color}

\begin{document}

\newcommand \be {\begin{equation}}
\newcommand \ee {\end{equation}}
\newcommand \bea {\begin{eqnarray}}
\newcommand \eea {\end{eqnarray}}
\newcommand \la {\langle}
\newcommand \ra {\rangle}
\newcommand \ve {\varepsilon}
\newcommand{\mC}{\mathcal{C}}
\newcommand{\mT}{\mathcal{T}}

\title[Statistical physics of interacting dissipative units]{Theoretical approaches to the steady-state statistical physics of interacting dissipative units}

\author{Eric Bertin}

\address{Universit\'e Grenoble Alpes and CNRS, LIPHY, F-38000 Grenoble, France}
\ead{eric.bertin@univ-grenoble-alpes.fr}

\begin{abstract}
The aim of this review is to provide a concise overview of some of the generic approaches that have been developed to deal with the statistical description of large systems of interacting dissipative 'units'.
The latter notion includes, e.g., inelastic grains,
active or self-propelled particles, bubbles in a foam,
low-dimensional dynamical systems like driven oscillators,
or even spatially extended modes like Fourier modes of the velocity field
in a fluid for instance.
We first review methods based on the statistical properties of a single unit, starting with elementary mean-field approximations, either static or dynamic, that
describe a unit embedded in a 'self-consistent' environment.
We then discuss how this basic mean-field approach can be extended to account for spatial dependences, in the form of space-dependent mean-field Fokker-Planck equations for example.
We also briefly review the use of kinetic theory in the framework of the Boltzmann equation, which is an appropriate description for dilute systems.
We then turn to descriptions in terms of the full $N$-body distribution, starting from exact solutions of one-dimensional models, using a Matrix Product Ansatz method when correlations are present.
Since exactly solvable models are scarce, we also present some approximation methods that can be used to determine the $N$-body distribution in a large system of dissipative units. These methods
include the Edwards approach for dense granular matter and the approximate treatment of multiparticle Langevin equations with coloured noise, which models systems of self-propelled particles.
Throughout this review, emphasis is put on methodological aspects of the statistical modeling and on formal similarities between different physical problems,
rather than on the specific behaviour of a given system.
\end{abstract}

\tableofcontents

\section{Introduction}

Equilibrium statistical physics fundamentally deals with the steady-state statistical properties of large assemblies of interacting conservative particles.
By conservative, one means here that the global energy of the system is conserved by the dynamics, and that a time-reversal symmetry holds.
In other words, the underlying microscopic dynamics of the system is Hamiltonian, thus providing symmetries and conservation laws that play a key role in the statistical description: the time-reversal symmetry underlies detailed balance, and the conservation of energy is at the root of the ensemble construction of equilibrium statistical physics ---see, e.g., \cite{Sokolov}.

Over the last decades, the domain of application of statistical physics has however progressively been extended to encompass different types of systems that are composed of dissipative particles.
The first example that has been extensively studied is probably the case of a granular gas, that is a gas of inelastic particles (e.g., steel or glass beads, or sand grains) that dissipate energy upon collisions \cite{Haff}. Kinetic theories, in the form of Boltzmann or Boltzmann-Enskog equations, have been quite naturally extended to this type of systems to describe their dilute or moderately dense regimes \cite{Poschel04}. Going to very dense regimes, granular matter has also raised important questions about how to describe the statistics of dense, mechanically stable, disordered packings of grains \cite{BHDC15}.

Another type of systems that has attracted a lot of attention more recently is that of assemblies of self-propelled (or more generally active) particles, modeling locally-driven colloids \cite{Cottin,Palacci} or more macroscopic particles \cite{Deseigne}, as well as small-scale biological systems powered by molecular motors, bacteria colonies and animal groups (mammal herds, fish schools, or bird flocks) \cite{Marchetti-RMP}.
Such active units are not conservative because they constantly dissipate energy, and are powered by (often chemical) energy taken from the environment.
Here also, different density regimes call for different methods. For dilute and moderately dense systems, kinetic theory methods \cite{EPJST14} and space-dependent mean-field approaches have been proposed \cite{Marchetti-RMP}.
In the opposite limit of high density, jammed systems of soft active particles may be a model of biological tissue \cite{Prost-PNAS}, and theoretical methods to study such jammed systems mostly remain to be developed.

Even in the absence of local activity, jammed packings of soft athermal particles are in themselves systems of interest; These may model soft disordered systems like foams, gels, or pastes. When applying an external shear, these otherwise blocked systems start to flow ---like when one presses on a toothpaste tube.
This flow actually results from localized cooperative rearrangements associated with plastic events \cite{Rodney11}. One may thus also think of considering these localized plastic events as the building blocks of a statistical description of such jammed systems, instead of considering the bubbles or the colloids as the elementary objects in the description. In any case, whether one takes a bubble or a plastic event as the elementary block, one is left with the description of interacting dissipative objects.

Many other examples can be found beyond the three examples above.
To cite only a few, one may think about
(i) coupled dynamical systems like in the Kuramoto model \cite{Acebron} and in models of coupled chaotic units \cite{ChateCML};
(ii) turbulent flows and simplified models like shell models, that can be thought of as a set of extended (typically Fourier) modes that interact through non-linear couplings \cite{Frisch,Bouchet-review,Ditlevsen};
(iii) gases of dissipative solitons as observed for instance in large scale simulations of self-propelled particles \cite{Solon15};
(iv) models of interacting socials agents, for which statistical physics approaches may be relevant in some cases \cite{Castellano,Bouchaud13,Jensen09,Jensen11}.

Reviews, or even textbooks, of course exist on most of these topics taken separately, see e.g., \cite{Poschel04} for granular gases, \cite{BHDC15} for dense granular packings, \cite{Marchetti-RMP} for active matter systems, \cite{Bouchet-review} for two-dimensional turbulence in fluids, or \cite{Castellano} for models of social agents.
However, taken together, all these examples suggest that a more global statistical physics approach of systems of interacting dissipative units would certainly be desirable.
It is of course not clear at this stage whether a single, unified framework could be built to deal with the many possible types of different systems that could enter this broad category.
Yet, one of the goals of this review is to suggest to think of all those systems in a common statistical physics perspective, since their statistical description raises common questions in many cases (dissipative interactions, absence of detailed balance, difficulties to determine the stationary phase-space distribution,...).

With this aim in mind, we have organized the review into four main sections corresponding to different types of methods that can be used to describe the statistics of a system of dissipative units. In this sense, this review is much more focused on methodological aspects than on the detailed study of specific systems.
Each section typically presents two or three models that can be studied with the method considered, and formal similarities between the problems described within a given section are emphasized.
Note that the presentation of each model does not go into the detailed behaviour of the system, but focuses on the derivation of the phase-space (or configuration space) probability distribution, either for one particle or for the $N$ particles composing the system, and when relevant on the derivation of more macroscopic information like hydrodynamic equations or averaged global observables.
In each case, the physical insights gained from the results are very briefly sketched.
Further, the order of the four sections has been chosen in order to go smoothly from simple mean-field approximation methods to more involved mean-field or kinetic theory methods that take into account spatial dependence, and finally to exact or approximate methods based on the determination of the full phase space distribution.

The article is organized as follows. 
Sect.~\ref{sec-MF} describes elementary mean-field methods that focus on a single unit and treat the rest of the system as a self-consistent environment.
Sect.~\ref{sec-MFFP} presents improved versions of this mean-field approach, that are able to retain spatial information.
Sect.~\ref{sec-kinetic-th} then describes simple kinetic theories based on the Boltzmann equation, and emphasizes the similarities and differences with the space-dependent mean-field approaches.
After these first sections that considered the statistical description of a single unit, we turn to methods aiming at determining the full $N$-body distribution of the system.
Sect.~\ref{sec-Nbody} starts by discussing exactly solvable models, with specific emphasis on one-dimensional models for which a solution can be found through a Matrix Product Ansatz.
The remainder of the section then introduces some approximations methods that can be used to determine the $N$-body distributions in more realistic models that cannot be solved exactly.
Finally, a summary and an outlook are provided in Sect.~\ref{sec-conclusion}.


\section{Mean-field approaches}
\label{sec-MF}

Generally speaking, mean-field approaches consist in reducing the description of the full system composed of a large number $N$ of interacting units to the description of a single unit subjected to effective average interactions. This can be done either by making (sometimes crude) approximations on the true interactions, or by considering a fully-connected version of the model in which all units are coupled together and interact in the same way (at odds with what happens, for instance, if the units are placed on a lattice and only interact with their neighbors).
In the fully-connected case, calculations can in most cases be performed exactly, but the approximation is in a sense in the definition of the model with respect to the physically relevant situation.
In the other cases, approximations consists in neglecting correlations in the system, for instance by assuming that all the degrees of freedom of the system are statistically independent.
In practice, many different forms of mean-field approximations exist, ranging from well-formalized ones (like assuming full statistical independence) to more phenomenological ones, using a purely adhoc modeling of the environment of a given unit.
Note that in many cases, mean-field approximations are convenient, but not well controlled. In addition, an adhoc modeling of the environment is actually more a way to model the system, partly based on physical intuition, than a systematic approximation scheme applied to the dynamics of the system. The other approximation methods proposed, however, are more systematic.

Another type of distinction between different mean-field approaches is whether they are static or dynamic. In the static case, one generally determines an approximation of the one-body stationary distribution,
for instance by maximizing the entropy of this distribution
under some constraints that can be evaluated in a self-consistent way from the one-body distribution.
No dynamics of the systems needs to be explicitly specified.
This approach is illustrated on two examples in Sect.~\ref{sect-static-MF}.
By contrast, the dynamic approach consists in defining first a (stochastic or deterministic) dynamics of the system, in which the effect of other units is treated either through some phenomenological approximations, or through a global coupling. 
The evolution of the one-body probability distribution is then generally governed by a non-linear partial differential equation, or a non-linear integro-differential equation.
Several examples of this approach are given in Sect.~\ref{sect-dyn-MF}.

\subsection{Static mean-field approximation}
\label{sect-static-MF}

As mentioned above, the general spirit of the static mean-field approximation is to focus on a single unit, and to make approximations on the static constraints imposed by the rest of the system, when evaluating the steady-state probability distribution of the configurations of this single unit.
Two examples of such a static mean-field approach are provided below:
on the one hand a simple model of a disordered packing of frictional grains, and on the other hand a model of two-dimensional foam.

\subsubsection{Disordered packing of frictional grains}
\label{sec-levine-grains}

As a first example, we consider the case of dense granular matter, that is, a dense assembly of macroscopic grains interacting via contact forces and dry friction.
In the absence of external driving, the system relaxes to a mechanically stable configuration under the effect of gravity.
Applying a strong permanent driving, for instance by shaking the container, would lead to a granular gas ---see Sect.~\ref{sec-granul-gas} below.
Here, however, we are interested in a more intermittent type of driving, like a tapping dynamics, in which the system periodically undergoes driven and undriven phases. In this situation, the system relaxes to a mechanically stable configuration before the next driving phase starts, and one may perform a statistics over the successively visited mechanically stable configurations.

In this section, we consider mean-field approaches that focus on a single particle. Different approaches of this kind have been put forward. It has in particular been argued that the local properties of the packing could be described 
from the statistics of the solid angle attached to each neighbor (seen from the focus grain) using random walk properties \cite{Clusel09}.
An alternative approach, that we describe in more detail below, is to follow
the general strategy proposed by Edwards and coworkers \cite{EO89,ME89,EM94,EG98} and later further developed by other groups \cite{BKVS01,LD03,BSWM08,HC09,WSJM11,BZBC13,Daniels,APF14}, and to proceed by analogy with equilibrium by assuming that all configurations compatible with the constraints are equiprobable. Before discussing how to implement this procedure in practice, two comments are in order. The first comment is that contrary to the equilibrium situation, the dynamics of the system is dissipative, and there is no underlying microreversibility property to back up the equiprobability assumption.
Second, the constraints to be taken into account are not only the global constraints on the conservation of the total volume or energy like at equilibrium,
but also a more complicated constraint accounting for mechanical stability.
Configurations that are not mechanically stable are assigned a zero probability weight.
An interesting property of dry friction, as opposed for instance to viscous friction, is that the set of stable configurations is in many cases of finite (i.e., nonzero) measure in the set of all possible configurations.
In contrast, viscous friction makes (in the absence of dry friction) the system relax to the local minima of the potential energy landscape, and this set of minima is in general of zero measure.

A standard way to implement the Edwards' prescription is to use a ``canonical'' version in which quantities like volume and energy are not fixed, but are allowed to fluctuate, with fluctuations described by Boltzmann-like factors. 
A general form of the probability of a given configuration $\mathcal{C}$ of the system is then
\be
P(\mathcal{C}) = \frac{1}{Z} \exp\left(-\frac{E(\mathcal{C})}{T_{\rm Edw}} 
-\frac{V(\mathcal{C})}{X} \right) \, \mathcal{F}(\mC),
\ee
where $Z$ is a normalization factor,
$E(\mathcal{C})$ and $V(\mathcal{C})$ are respectively the energy and volume associated with configuration $\mathcal{C}$,
$T_{\rm Edw}$ is an effective temperature, and $X$ is an effective ``thermodynamic'' parameter called compactivity. In equilibrium systems, $X^{-1}$ would be equal to the ratio $P/T$ of pressure and temperature (note that throughout this review, we set the Boltzmann constant $k_B=1$).
In addition, the function $\mathcal{F}(\mC)$ restricts the probability measure to configurations that are mechanically stable: $\mathcal{F}(\mC)=1$ if $\mC$ is mechanically stable, and $\mathcal{F}(\mC)=0$ otherwise.
In more detailed presentation of Edwards' statistical mechanics is provided in Sect.~\ref{sec:Edwards-Nbody}.

\begin{figure}[t!]
\centering\includegraphics[width=7cm]{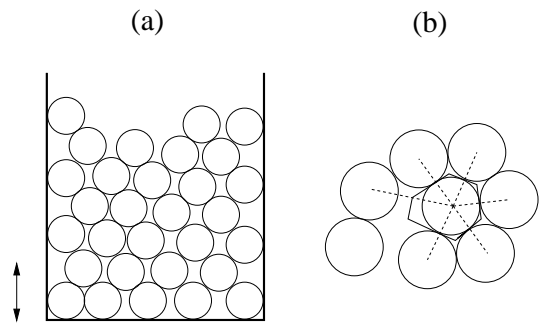}
\caption{(a) Sketch of the granular packing, enclosed in a contained periodically subjected to vertical vibration phases. (b) Schematic representation of the Voronoi cell surrounding a given grain.}
\label{fig-granul-pile}
\end{figure}

\paragraph{Mean-field distribution of the local volume}

In the following, we neglect the influence of energy, and focus on the effect of the volume constraint. To determine the volume $V(\mathcal{C})$ of configuration $\mathcal{C}$, it is then convenient to assign a volume to each grain using a Voronoi tesselation (see Fig.~\ref{fig-granul-pile}) \cite{Levine03,Dauchot06}:
\be
V(\mathcal{C}) = \sum_{i=1}^N v_i(\mathcal{C})
\ee
with $N$ the total number of grains, and $v_i(\mathcal{C})$ the volume of the Voronoi cell around grain $i$ (hereafter simply denoted as cell $i$). Since the individual volume $v_i(\mathcal{C})$ also depends on the position of the neighbouring grains, and not only on the position of grain $i$, the individual volumes are correlated.
In principle, it is possible to obtain the marginal distribution $P_i(v)$ of the volume of cell $i$ by summing over all configurations $\mathcal{C}$ such that $v_i(\mathcal{C})=v$. In practice, this calculation is not easily tractable, because the function $v_i(\mathcal{C})$ takes a complicated form.
To simplify the problem, a mean-field approximation consists in
assuming that the volumes of the Voronoi cells are statistically independent, with identical distributions.
Furthermore, the constraint of mechanical stability encoded in the complex function $\mathcal{F}(\mC)$ may be simply expressed, at mean-field level, as
a maximal accessible local volume $v_{\rm max}$ (in addition to the minimal local volume $v_{\rm min}$ of purely steric origin).
Under these assumptions, one obtains that the distribution $P(v)$ of the individual volume takes the form
\be \label{eq:dist-PvgranMF}
P(v) = \frac{1}{Z_1}\, g(v) \, e^{-v/X}, \qquad v_{\rm min} < v < v_{\rm max},
\ee
where $Z_1$ is a normalization constant and $g(v)$ is the (unknown) density of states.
Since there is no simple way to determine the density of states without a more detailed modeling of the packing, we further simplify the problem by assuming, following \cite{Levine03}, that the density of states is uniform over the interval $v_{\rm min} < v < v_{\rm max}$.
In practice, we thus use Eq.~(\ref{eq:dist-PvgranMF}) with $g(v)=1$.

The minimal volume $v_{\rm min}$ corresponds to the densest local packing, and depends on the shape of the grains; It is determined only by steric constraints.
For circular disks of radius $r_0$ in 2D, the most compact packing is the hexagonal packing, yielding $v_{\rm min}^{2D} = \sqrt{12} r_0^2$.
In 3D, it is given by $v_{\rm min}^{3D} = \sqrt{32} r_0^3$, which is achieved for the face centered cubic or the hexagonal close packing \cite{Levine03}.
The determination of the maximal volume $v_{\rm max}$ is more interesting, and requires more input from the physics of the problem. A key ingredient here is the presence of dry friction, which allows the grains to support some tangential forces, and tends to stabilize less dense configurations. As a result, one expects the maximal volume $v_{\rm max}$ to be an increasing function of the static friction coefficient $\mu_{\rm s}$.
Approximate expressions of $v_{\rm max}$ as a function of $\mu_{\rm s}$ can be derived, assuming very simple geometries \cite{Levine03}.
Considering now that $v_{\rm max}$ is known,
one can determine from Eq.~(\ref{eq:dist-PvgranMF}) the average value of $v$, as well as higher order moments. One finds for the average value \cite{Levine03,EO89}
\be
\la v \ra = \frac{1}{2}(v_{\rm min}+v_{\rm max}) + X - \Delta v \coth \left( \frac{\Delta v}{X} \right),
\ee
where $\Delta v=(v_{\rm max}-v_{\rm min})/2$.
It is then possible to determine the volume fraction $\Phi$ as a function of $\mu_{\rm s}$, defining $\Phi=v_0/\la v \ra$, where $v_0$ is the volume of a grain.

\paragraph{Segregation phenomenon}

Interestingly, this simple model can be generalized to qualitatively describe the segregation of grains having different frictional properties. 
Considering an assembly of monodisperse grains of two different types A and B, one can define the friction coefficients $\mu_{AA}$, $\mu_{AB}$ and $\mu_{BB}$ corresponding to A-A, A-B and B-B contacts respectively.
As a further simplification, one may then consider only three distinct maximal volumes $v_{AA}$, $v_{BB}$ and $v_{AB}=v_{BA}$, where $v_{jk}$ corresponds to the maximal accessible volume for the Voronoi cell of a grain of type $j$ surrounded by grains of type $k$ ($j,k=A,B$).
Denoting as $f$ the fraction of $A$ grains, one has in a mean-field approximation that the numbers $n_{jk}$ of grains $j$ surrounded by grains $k$ are
$n_{AA}=f^2 N$, $n_{AB}=n_{BA}=f(1-f)N$ and $n_{BB}=(1-f)^2 N$.
The different volumes $v_i$ being assumed statistically independent, 
one can write the $N$-particle partition function\footnote{We have characterized above the mean-field approach as the description of a single focus particle, making approximations to describe the environment of this particle. Once these approximations are done, one can also equivalently describe the system as a set of $N$ independent particles, which may be more convenient in some situations like the present one.} as \cite{Levine03}
\bea \nonumber
Z_N &=& \frac{N!}{(fN)! [(1-f)N]!}
\left( \int_{v_{\rm min}}^{v_{AA}} e^{-v/X} dv \right)^{f^2 N}\\
&& \qquad \times
\left( \int_{v_{\rm min}}^{v_{AB}} e^{-v/X} dv \right)^{2f(1-f)N}
\left( \int_{v_{\rm min}}^{v_{BB}} e^{-v/X} dv \right)^{(1-f)^2 N} .
\eea
Note that as grains are monodisperse, the volume $v_{\rm min}$ is the same for all grains.
One can then define the analogue of a free energy as $Y=-N^{-1}X\ln Z$, yielding
\bea \nonumber
Y &=& X[f\ln f + (1-f)\ln(1-f) + 2f(1-f) R(X) \\
&& \qquad \qquad \qquad \qquad  -\ln X -f R_{AA}(X)-(1-f) R_{BB}(X)]
\eea
with the notations
\bea
R_{jk}(X) &\equiv& \ln(e^{-v_{\rm min}/X}-e^{-v_{jk}/X}), \\
R(X) &\equiv& \frac{1}{2}[R_{AA}(X)+R_{BB}(X)]-R_{AB}(X).
\eea
The function $Y$ can be used to determine whether a phase separation (i.e., a segregation) occurs or not by minimizing $Y$ as a function of $f$ under the constraint of fixed average fraction $\bar f$. 
In the case $\bar f=\frac{1}{2}$, the fraction $f$ is then found to satisfy
\be
2f-1 = \tanh [R(X) (2f-1)],
\ee
an equation similar to the self-consistent equation obtained when solving the mean-field Ising model.
For $R(X)<1$, the only solution is $f=\frac{1}{2}$, and the system remains homogeneous.
In constrast, when $R(X)>1$, two solutions $f \ne \frac{1}{2}$ exist on top of the solution $f=\frac{1}{2}$ which is no longer the stable one; As a result,
segregation is obtained.
Whether $R(X)$ can be larger than $1$ in some range of compactivity $X$ actually depends on the values of the friction coefficients.
In the case $\mu_{AB}=\min(\mu_{AA},\mu_{BB})$, one finds that a segregated state exists ($R(X)>1$) in a range of $X$ when $\mu_{AA}$ and $\mu_{BB}$ take sufficiently distinct values.
Using the relation between compactivity and volume fraction, it is also possible to characterize the segregation phenomenon in terms of volume fraction \cite{Levine03}.

\subsubsection{A model of two-dimensional foam}
\label{sec-foam}

A relatively similar statistical approach can be performed in the case of a two-dimensional foam \cite{Iglesias91,Graner00,Durand10,Durand11,Durand14,Durand15}.
The idea is to describe the statistics of the configurations of a foam that are visited due to a slow externally applied shear rate. The number $N_B$ of bubbles is assumed to be constant on the involved time scales, and rearrangements are assumed to occur through so-called 'T1' processes ---the most elementary topological rearrangement.
The elementary objects in the description of the foam are the bubbles, considered to have a fixed area $A_i$ (which corresponds to having a fixed volume of the bubbles in an experiment confined in the third dimension).
Contrary to the granular system, there is no exchange of area (i.e., of volume) between bubbles. Rather, the fluctuating variable at the bubble scale is the number $n_i$ of sides of each bubble $i$.
A configuration of the system is thus defined as $(n_1,\dots,n_N)$.

\begin{figure}[t!]
\centering\includegraphics[width=5cm]{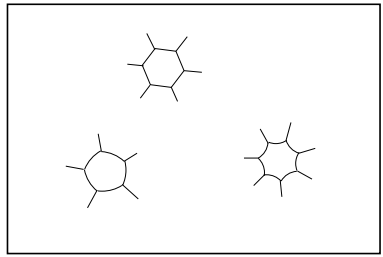}
\caption{Sketch of a foam in the mean-field picture, where ideal bubbles with $n$ sides (illustrated here for $n=5$, $6$ and $7$) are singled out from their environment.}
\label{fig-foam}
\end{figure}

In order to determine the statistics of the configuration $(n_1,\dots,n_N)$,
one first has to identify the constraints to which the system is subjected.
A first constraint is that for a large system, the average number of sides is equal to $6$,
\be \label{eq:bb:constraint1}
\frac{1}{N_B} \sum_{i=1}^{N_B} n_i = 6 .
\ee
A second constraint is related to the Laplace law, which relates the algebraic curvature $\kappa_{ij}$ of the side joining bubbles $i$ and $j$ to the difference of pressure between these two bubbles:
\be \label{eq:Laplace:law}
\kappa_{ij} = \frac{P_j-P_i}{\gamma}
\ee
where $P_i$ is the pressure in bubble $i$, and $\gamma$ is the film tension.
Note that $\kappa_{ij}$ is the curvature of the side joining bubbles $i$ and $j$ when the side is considered as belonging to bubble $i$.
If the side is considered from the point of view of bubble $j$, its curvature is $\kappa_{ji} =-\kappa_{ij}$.
Defining the total curvature of a bubble as
\be
\kappa_i^{\rm tot} = \sum_{j \in V(i)} \kappa_{ij}
\ee
where $V(i)$ is the set of neighboring bubbles of bubble $i$,
one obtains from the Laplace law (\ref{eq:Laplace:law}) that
\be \label{eq:bb:constraint2}
\frac{1}{N_B} \sum_{i=1}^{N_B} \kappa_i^{\rm tot} = 0,
\ee
which is the second global constraint on the system.
The first constraint, Eq.~(\ref{eq:bb:constraint1}), is easy to take into account as it depends explicitly on the numbers $n_i$ of sides.
This is not the case of the second constraint Eq.~(\ref{eq:bb:constraint2}).
Hence the effective dependence of the total curvature $\kappa_i^{\rm tot}$
on the number $n_i$ of sides has to be modeled, which can be done
using the following mean-field argument.
Neglecting correlations with the surrounding bubbles, one may simply consider
a bubble as a regular cell with $n$ identical curved sides, forming an angle of $120^{\circ}$ at each vertex, where two sides of the bubble considered join with a third side separating neighbouring bubbles (see Fig.~\ref{fig-foam}).
In this case, the Gauss-Bonnet theorem states that (see, e.g., \cite{Durand15})
\be \label{Gauss-Bonnet}
n\frac{\pi}{3}-n\kappa_0 \ell = 2\pi
\ee
where $\kappa_0$ and $\ell$ are respectively the curvature and length of each side of the cell.
Eq.~(\ref{Gauss-Bonnet}) simply states that the angular variations of the tangent vector sum up to $2\pi$ when going around the cell.
The total curvature $\kappa^{\rm tot}=n\kappa_0$ is then obtained from Eq.~(\ref{Gauss-Bonnet} as
\be
\kappa^{\rm tot} = \frac{\pi}{3} \frac{n(n-6)}{P}
\ee
where $P=n\ell$ is the perimeter of the cell. Since bubbles have fixed area $A$, it is convenient to express the perimeter $P$ as a function of the area $A$ and of the number $n$ of sides. For dimensional reasons, one has $P=e(n)\sqrt{A}$;
The quantity $e(n)$ is called the elongation of the cell. It turns out that $e(n)$ remains very close to a constant value $\approx 3.72$ (the value corresponding to a hexagon) over the relevant range of values of $n$ \cite{Graner00,Durand10}. As a result, we end up with the simple expression
\be
\kappa^{\rm tot} \approx \kappa(n,A) \equiv c \, \frac{n (n-6)}{\sqrt{A}}
\ee
where $c \approx 0.28$ is a constant.

In the following, we focus on the determination of the distribution $p_A(n)$ of the number of sides $n$ for a given value of the area $A$.
In the mean-field framework considered here, 
the distribution $p_A(n)$ is then obtained by maximizing the entropy
$S=-\sum_n p_A(n) \ln p_A(n)$ of the one-body distribution under the constraint that
the average values $\la n \ra$ and $\la \kappa(n,A) \ra$ are fixed.\footnote{At equilibrium, the maximization of entropy can be justified by the underlying time-reversibility of the microscopic dynamics. In dissipative systems, no such justification exists. The meaning of entropy maximization is rather that of a maximum likelihood principle: in the absence of further information (like the knowledge of a dynamics), the most likely distribution is the one that maximizes entropy under the known constraints.} This procedure leads to
\be \label{eq:pn:bb}
p_A(n) = \frac{1}{Z} \, e^{-\beta \kappa(n,A) -\mu n}
\ee
where $Z=\sum_n e^{-\beta \kappa(n,A) -\mu n}$ is a normalization constant,
and $\beta$ and $\mu$ are Lagrange parameters.
An alternative interpretation is that the rest of the system acts as a reservoir of both sides and curvature, so that Eq.~(\ref{eq:pn:bb}) corresponds to a
grand-canonical distribution with ``thermodynamic'' parameters $\beta$ and $\mu$. 

One of the interests of the distribution (\ref{eq:pn:bb}) is that it predicts a correlation between the topology of bubbles (their number $n$ of sides) and their geometry (their area $A$). This prediction can be tested against experimental data on foams in several ways.
The distribution $p_A(n)$ given in Eq.~(\ref{eq:pn:bb}) is defined for a fixed value of the area $A$. 
However, it is possible to average $p(n)$ over the statistics of bubble area, if the latter is known.
Taking the area distribution from experimental data, the average distribution ${\bar p}(n)$ of the number of sides has been successfully compared to the experimentally measured distribution $p_{\rm exp}(n)$ \cite{Durand10}.
Moreover, a linear relation between the average number of sides $\langle n \rangle_A$ as a function of the area $A$ has also been reported from experimental data
\cite{Durand11}. This relation can be understood from a Gaussian approximation of the distribution $p_A(n)$ given in Eq.~(\ref{eq:pn:bb}) \cite{Durand11}.

\subsection{Dynamic mean-field approximation}
\label{sect-dyn-MF}


In the dynamic mean-field approach, approximations (or assumptions of global couplings) are rather made at the level of the dynamics. The steady-state probability distribution of the single unit considered has to be determined from the evolution equation of the probability distribution, typically a master equation or a Fokker-Planck equation.
The price to pay for the simplification of describing the evolution of a single unit instead of the whole system is that the evolution equation for the probability distribution is non-linear (and, of course, only provides an approximate description of the system).
This dynamic mean-field approach is illustrated below on three different examples.
Sect.~\ref{sec-HL} presents a model describing plastic rearrangements in a driven elastoplastic system. In this model, the mean-field approximation is implemented as a phenomenological description of the rest of the system on the focus unit. Then Sect.~\ref{sec-granul-gas} and \ref{sec-Kuramoto}
respectively describe a stochastic granular gas model and a deterministic model of coupled driven oscillators. Both models are defined with global couplings, so that they can be solved exactly without further approximation.

\subsubsection{H\'ebraud-Lequeux model for sheared elastoplastic systems}
\label{sec-HL}


Let us start by discussing an example where the influence of the environment of the focus unit is described in a phenomelogical way. We consider the H\'ebraud-Lequeux model \cite{HL98,Agoritsas15,Bouchaud16}, which describes the statistics of the local shear stress $\sigma$ in an elastoplastic model subjected to an imposed external strain rate $\dot\gamma$.
This type of model aims at describing the rearrangement dynamics under shear in soft materials like foams or assemblies of soft particles at high density, in the jammed state.
The dynamics of such soft glassy materials has also been described by models like the Soft Glassy Rheology (SGR) model \cite{SGR1,SGR2}, in which the effect of the environment is encoded in an effective temperature. However, the H\'ebraud-Lequeux model is more satisfactory at the conceptual level, since the noise is treated as a mechanical noise instead of an effective thermal noise (see, e.g., the discussion of this point in Refs.~\cite{Agoritsas15,Nicolas14}), and the intensity of the mechanical noise is determined in a self-consistent way, while it is just a model parameter in the SGR model.
Besides, note also that, although one might expect some links between the H\'ebraud-Lequeux model and the mean-field model of foams discussed in Sect.~\ref{sec-foam}, both models are actually unrelated: the H\'ebraud-Lequeux model is not specific to foams, and its focus is on internal stresses rather than on the geometry of bubbles.

The rearrangement dynamics is known to be heterogeneous in this case and to proceed via elastic loading followed by localized plastic events during which local stress is released, and redistributed over the whole system via a long-range elastic propagator.
The basic idea is to decompose the system into mesoscopic cells of the size of the locally rearranging regions.
Each cell can be in an elastic or plastic state, and the transition to a plastic state occurs when the local stress, which increases in the elastic state due to the applied deformation rate $\dot\gamma$, exceeds a threshold value.
In the plastic state, the local stress relaxes until the elastic state is reached again. Stress is redistributed to other cells through the long-range elastic propagator during plastic relaxation.
The resulting lattice model has been studied numerically \cite{Picard04,Picard05,Martens11,Martens12,Martens14}, or using mean-field approximations to get analytical results \cite{HL98,Agoritsas15,Bouchaud16}.

\begin{figure}[t!]
\centering\includegraphics[width=5cm]{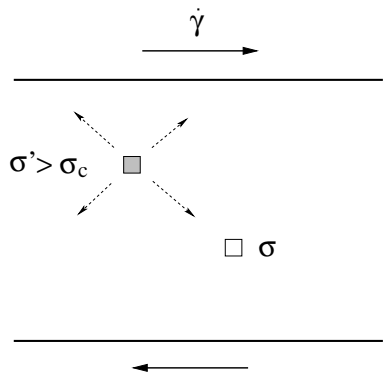}
\caption{Sketch of the H\'ebraud-Lequeux model. The empty square represents the focus cell, with local stress $\sigma$. Under the effect of the external shear rate $\dot{gamma}$, another cell (represented in grey) with stress $\sigma'>\sigma$, undergoes a plastic rearrangement. The resulting stress drop propagates elastically within the system, and may contribute to trigger a plastic event on the focus cell.}
\label{fig-HL}
\end{figure}

\paragraph{Mean-field equation for the one-site distribution}
In a mean-field description, one focuses on the evolution of a single cell, and considers the remaining cells as simply generating a mechanical noise
(see Fig.~\ref{fig-HL}).
The mean-field assumption thus neglects the correlation of the focus cell with its surrounding, as well as the potentially complex temporal structure of the stress signal received from the rest of the system, simply considering it as a white noise.
The local stress $\sigma$ is thus subject to both a drift dynamics
with 'velocity' $\mu \dot\gamma$ ($\mu$ being the elastic shear modulus) resulting from the externally applied shear, and a diffusion dynamics with diffusion coefficient $D(t)$ corresponding to the mean-field description of plastic events occuring throughout the system.
In addition, plastic events in the cell considered
are assumed to occur randomly, with a probability $1/\tau$ per unit time, when the local stress $\sigma$ exceeds a threshold value $\sigma_c$.
For simplicity, the subsequent plastic relaxation is then considered as both instantaneous and complete, meaning that at the end of the relaxation, $\sigma=0$.

Due to the stochastic nature of plastic events, the local stress $\sigma$ is a random variable, whose probability distribution $P(\sigma,t)$ evolves according to
\be \label{eq:dyn-HL}
\frac{\partial P}{\partial t} = -\mu \dot\gamma \frac{\partial P}{\partial \sigma} + D(t) \frac{\partial^2 P}{\partial \sigma^2} - \frac{1}{\tau} \Theta(|\sigma| -\sigma_c) + \Gamma(t) \delta(\sigma),
\ee
where $\Theta$ is the Heaviside step function, $D(t)$ is the effective diffusion coefficient resulting from plastic relaxations of other cells,
and $\Gamma(t)$ is the plastic activity defined as
\be
\Gamma(t) = \frac{1}{\tau} \int_{|\sigma|>\sigma_c} P(\sigma,t) \, d\sigma \,.
\ee
Physically, the diffusion coefficient $D(t)$ is expected to depend on the plastic activity $\Gamma(t)$: the more plastic events are present, the larger the diffusion coefficient. The H\'ebraud-Lequeux model simply assumes a linear relation
\be \label{eq:DGamma}
D(t) = \alpha \Gamma(t),
\ee
where $\alpha$ is a parameter of the model.
Hence the diffusion coefficient $D(t)$ has to be determined self-consistently from the stress distribution $P(\sigma,t)$, through the plastic activity $\Gamma(t)$. The self-consistency relation (\ref{eq:DGamma}) thus makes the evolution equation (\ref{eq:dyn-HL}) non linear.

Note that in this model, the effect of the rest of the system is treated on a phenomenological basis. Another possibility could be to start from the description of the full system, and to integrate over all cells but one, possibly under simplifying assumptions. This is the strategy followed by the so-called Kinetic Elasto-Plastic (KEP) model, that we will briefly describe in Sect.~\ref{sec-KEP}.
In this way, one would derive Eq.~(\ref{eq:dyn-HL}) from the dynamics of the full system.

Note also that Eq.~(\ref{eq:dyn-HL}) is reminiscent of the stochastic process called ``diffusion with stochastic resetting'' \cite{Evans2011}.
However, the H\'ebraud-Lequeux model differs from the latter process because
it includes a non-zero drift term, and more importantly because the diffusion constant is not fixed, but determined self-consistently.
The H\'ebraud-Lequeux model may thus be thought of as a generalization of the diffusion with stochastic resetting process.

\paragraph{Determination of the average stress}
The stationary solution of Eq.~(\ref{eq:dyn-HL}) can be worked out exactly, at least in an expansion in powers of the strain rate $\dot\gamma$, in the low strain rate limit \cite{HL98,Agoritsas15}.
Technically, this is done by first considering $D$ as a given parameter, and solving the equation over the intervals $(-\infty,-\sigma_c)$, $(-\sigma_c,0)$, $(0,\sigma_c)$ and $(\sigma_c,\infty)$.
The distribution $P(\sigma)$ is then obtained by taking into account matching conditions at the interval boundaries, as well as the self-consistency relation $D=\alpha \Gamma$, where $\Gamma$ can be evaluated as a function of $D$ from the solution $P(\sigma)$ obtained at fixed $D$.
The resulting stationary distribution $P(\sigma)$ does not, however, take a simple form \cite{Agoritsas15}, and we thus do not report it here explicitly.

The physically important quantity is the average stress $\la \sigma \ra$, expressed as a function of $\dot\gamma$ (this relation is called the rheological law).
One finds that for $\alpha>\alpha_c \equiv \frac{1}{2}\sigma_c^2$, the average stress goes to zero when $\dot\gamma \to 0$, while for $\alpha>\alpha_c$ a finite yield stress $\sigma_Y>0$ emerges
\be \label{eq:HB}
\la \sigma \ra \approx \sigma_Y + A \dot\gamma^{1/2}, \qquad \dot\gamma \to 0.
\ee
Such a behaviour is called a Herschel-Bulkley law in the rheological literature.
Both the yield stress $\sigma_Y$ and the prefactor $A$ depend on $\alpha$, and can be explicitly determined.
For $\alpha \lesssim \alpha_c$, $\sigma_Y$ is given by
\be
\sigma_Y \sim (\alpha_c-\alpha)^{1/2} \,.
\ee
Interestingly, for $\alpha = \alpha_c$, a non-trivial scaling relation
$\la \sigma \ra \sim \dot\gamma^{1/5}$ is obtained.
Note that Eq.~(\ref{eq:HB}), which can be reformulated as $\dot\gamma \sim (\la \sigma \ra - \sigma_Y)^2$, suggests the existence of an underlying critical point at $\dot\gamma=0$ \cite{Rosso}.

To sum up, the interest of the H\'ebraud-Lequeux model is two-fold.
First, it provides a theoretical description, starting from a mesoscopic dynamics, of the Herschel-Bulkley law (\ref{eq:HB})
which is observed experimentally in the rheology of complex fluids
\cite{Hohler,Manneville}.
Second, as already mentioned in the introduction of the present subsection,
this model proposes a conceptually satisfying scenario for athermal systems, since mechanical noise is self-generated by the dynamics, while other models rather characterize mechanical noise through a fixed effective temperature
\cite{SGR1,SGR2}, in close analogy to thermal systems.

\subsubsection{Granular gas model}
\label{sec-granul-gas}

Having discussed in the previous subsection how the environment of the focus unit can be modeled in a phenomenological way, we now discuss two examples of a more systematic approach to model this environment, using models with global interactions.
In the present subsection, we consider a fully-connected granular gas model; The case of a model of globally-coupled driven oscillators is discussed in Sec.~\ref{sec-Kuramoto}.

When strongly shaken, an assembly of inelastic grains enclosed in a container enters a granular gas state, in which the density is roughly homogeneous, and energy is injected by collisions with the container and dissipated through binary collisions between grains --- see, e.g., \cite{Poschel04,Sela95,Shapiro95,Brey97,Trizac99,Trizac02,Trizac06,Gradenigo11}.
However, the presence of boundary injection leads to spatial heterogeneities: grains close to the boundaries have on average more kinetic energy.
Although boundary driving is the experimentally relevant situation, one may,
from a theoretical perspective, try to define a model which remains spatially homogeneous
(at least in the absence of any instability of the homogeneous state).
Such a model can be obtained by randomly injecting energy to all particles, irrespective of their position.
In such a model, a given particle may collide with any other particle, but the dynamics generates correlations between particles. There is also typically a higher probability to collide with a fast particle than with a slow one.

\begin{figure}[t!]
\centering\includegraphics[width=5cm]{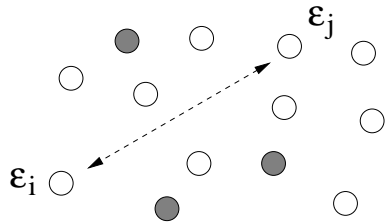}
\caption{Schematic illustration of the mean-field granular gas.
Any pair of particles can exchange energy through a dissipative interaction, irrespective of their position. Bath particles (in grey) have an energy drawn from a Boltzmann distribution with temperature $T_B$, and play the role of an energy injection mechanism.}
\label{fig-granul-gas}
\end{figure}

In a mean-field approach, one may simplify the problem by neglecting all correlations, and assume that any pair of particles may collide with a given probability, which does not depend on the configuration of the system
(see Fig.~\ref{fig-granul-gas}).
A simplified model of this type has been studied in \cite{Levine04}.
This model considers a large set of particles and focuses on the dynamics of a given particle. This focus particle with energy $\ve_i$ undergoes at rate $(1-f)\Gamma$ binary inelastic collisions with other randomly chosen particles having energy $\ve_j$; $\Gamma$ is a frequency scale, and $0 < f \le 1$ is a parameter of the model characterizing the relative frequency of energy dissipation and injection.
After a collision, the energy of the focus particle $i$
is assumed to be given by
\be
\ve_i' = z\alpha (\ve_i+\ve_j)
\ee
where $0\le \alpha \le 1$ is the inelasticity coefficient, and $0\le z \le 1$ is a uniformly distributed random variable determining the share of the remaining energy that is assigned to particle $i$ (a non-uniform distribution of $z$ could also be considered).
In addition, the particle also undergoes, with rate $f\Gamma$, elastic collisions with particles of a heat bath at temperature $T_B$. Particles from the bath have an energy $\ve_B$ distributed according to $p_B(\ve_B)=T_B^{-1} \exp(-\ve_B/T_B)$. After a collision with a bath particle, the focus particle has an energy
$\ve_i' = z (\ve_i+\ve_B)$.

The probability distribution of the energy $\ve$ of the focus particle at time $t$ is denoted as $p(\ve,t)$.
Its evolution is described by a self-consistent non-linear master equation
with time-dependent transition rates $W_t(\ve'|\ve)$ given by
\bea \nonumber
W_t(\ve'|\ve) &=&
(1-f) \Gamma \int_0^\infty d\ve'' p(\ve'',t) \int_0^1 dz \,
\delta \big( z\alpha(\ve+\ve'')-\ve'\big) \\
&& \qquad \qquad + f\Gamma \int_0^\infty d\ve_B\, p_B(\ve_B) \int_0^1 dz \,
\delta \big( z(\ve+\ve_B)-\ve'\big).
\label{eq:tr-rates:grangas}
\eea
The first term in the rhs of Eq.~(\ref{eq:tr-rates:grangas})
describes inelastic collisions between two particles of the granular gas, while the second term describes elastic collisions with particles of the bath.
The corresponding master equation reads
\bea \nonumber
\frac{\partial p}{\partial t}(\ve,t) &=& -\Gamma p(\ve,t)\\ \nonumber
&+& (1-f) \Gamma \int_0^\infty d\ve' p(\ve',t)
\int_0^\infty d\ve'' p(\ve'',t) \int_0^1 dz \,
\delta \big( z\alpha(\ve'+\ve'')-\ve \big) \\
&+& f \Gamma \int_0^\infty d\ve' p(\ve',t)
\int_0^\infty d\ve_B \, p_B(\ve_B) \int_0^1 dz \,
\delta \big( z(\ve'+\ve_B)-\ve \big).
\label{eq:master:grangas}
\eea
To solve for the steady-state distribution $p(\ve)$, it is convenient to introduce the Laplace transform $g(\lambda)$ of $p(\ve)$, defined as
\be
g(\lambda) \equiv \int_0^{\infty} p(\ve)\, e^{-\lambda\ve} d\ve \, .
\ee
The time-independent version of Eq.~(\ref{eq:master:grangas}) then transforms into \cite{Levine04}
\be \label{eq:genfn:grangas}
g(\lambda) = (1-f) \int_0^1 dz \, g(\lambda z\alpha)^2
+f \int_0^1 dz \, \frac{g(\lambda z)}{1+\lambda z T_B}
\ee
where the explicit form of the distribution $p_B(\ve_B)$ has been taken into account, yielding for its Laplace transform $g_B(\lambda)=1/(1+\lambda T_B)$.
Eq.~(\ref{eq:genfn:grangas}) can be further transformed into a differential equation by making the change of variable $u=\lambda z$ in the two integrals, and then multiplying by $\lambda$ and differentiating with respect to $\lambda$.
One finally obtains the following equation \cite{Levine04}
\be \label{eq:genfn:grangas2}
\lambda g'(\lambda) = (1-f) g(\lambda \alpha)^2 +
\left( \frac{f}{1+\lambda T_B} -1 \right) g(\lambda) \,,
\ee
where $g'(\lambda)$ is the derivative of $g(\lambda)$.
Eq.~(\ref{eq:genfn:grangas2}) is supplemented by the condition $g(0)=1$,
which comes from the normalization of the distribution $p(\ve)$.
Eq.~(\ref{eq:genfn:grangas2}) can be analytically solved in limit cases.
It is easy to check that the equilibrium distribution at temperature $T_B$ is recovered both when $\alpha=1$ (elastic collisions) and when $f=1$ (no collisions between particles of the gas).
In the maximally dissipative case $\alpha=0$, one obtains an ordinary first order differential equation
\be
\lambda g'(\lambda) = \left( \frac{f}{1+\lambda T_B} -1 \right) g(\lambda)
+(1-f)
\ee
whose solution reads \cite{Levine04}
\be
g(\lambda) = \, _2 F_1(1,2,2-f,-\lambda T_B) (1+\lambda T_B)
\ee
where $_2 F_1$ is the Gauss hypergeometric function.
More generally, for $0<\alpha<1$ and $0<f<1$, Eq.~(\ref{eq:genfn:grangas2}) can be integrated numerically starting from $\lambda=0$.
Alternatively, by taking successive derivatives of Eq.~(\ref{eq:genfn:grangas2}) at $\lambda=0$, one can compute analytically the moments $\la \ve^n\ra$ of the distribution $p(\ve)$, using the relation
\be
\la \ve^n\ra = (-1)^n \frac{d^n g}{d\lambda^n}(0).
\ee
The knowledge of all the moments is equivalent to the knowledge of the distribution $p(\ve)$.

Although the resulting form of the distribution $p(\ve)$ is only known through its moments (or equivalently, through the generating function $g(\lambda)$),
it is clear that $p(\ve)$ shows significant deviations from an equilibrium distribution. For instance, it has been shown that a fluctuation-dissipation relation holds, but with a temperature different from that deduced from the average energy \cite{Levine04}; At equilibrium, both temperatures would coincide.

\subsubsection{Kuramoto model of coupled oscillators}
\label{sec-Kuramoto}

In this subsection, we provide another example of a dynamical model with a global coupling, this time with a deterministic dynamics. The interest of this second example is to illustrate a different type of solution method, well-suited for models where a phase transition occurs.

\paragraph{Overdamped limit of driven interacting oscillators}
The Kuramoto model is composed of $N$ globally coupled oscillators of phase $\theta_j$, with quenched random frequencies $\omega_j$.
To better understand the connection of this model to non-equilibrium systems, one can start, following \cite{Ruffo14}, from a model of driven coupled oscillators defined by the dynamics
\be \label{eq:langevin:kuram}
m\frac{d^2 \theta_j}{dt^2} = -\gamma \frac{d \theta_j}{dt}
+ \sum_{k=1}^N \tilde{K}_{jk} \sin(\theta_k-\theta_j) + \gamma \omega_j + \eta_j(t), \qquad j=1,\ldots, N
\ee
where $m$ is the 'mass' (or moment of inertia) of the oscillators,
$\gamma$ is the friction coefficient,
$\tilde{K}_{jk}$ is the coupling constant between oscillators $j$ and $k$,
and $\gamma \omega_j$ is the driving 'force' (or torque) acting on oscillator $j$ ($\omega_j$ has the dimension of a frequency);
$\eta_j(t)$ is a white noise satisfying
$\la \eta_j(t) \ra =0$ and $\la \eta_j(t) \eta_j(t') \ra = 2D\delta(t-t')$.
The equilibrium case corresponds to having both $\omega_j=0$ for all $j$
and symmetric couplings $\tilde{K}_{jk}=\tilde{K}_{kj}$.
In this case, Eq.~(\ref{eq:langevin:kuram}) describes a set of interacting oscillators coupled to a thermostat of temperature $T=D/\gamma$, with a Hamiltonian
\be
H = \sum_j \frac{p_j^2}{2m} + \frac{1}{2} \sum_{j \ne k} \tilde{K}_{jk} \cos(\theta_k-\theta_j),
\ee
with $p_j \equiv md\theta_j/dt$ the momentum conjugated to $\theta_j$.
In the presence of the driving force $\gamma \omega_j$ and possibly of non-symmetric coupling constants $\tilde{K}_{jk}$, the system becomes a set of interacting dissipative oscillators, and its statistics is no longer described by a Boltzmann-Gibbs probability weight.
The standard Kuramoto model \cite{Kuramoto75,Kuramoto84,Acebron} considers a specific limit of Eq.~(\ref{eq:langevin:kuram}), namely the overdamped limit at zero temperature.
The overdamped limit is obtained by taking the limit $m \to 0$ at a fixed (nonzero) value of the friction coefficient $\gamma$.
For $T=0$, one ends up with the evolution equation of the standard Kuramoto model \cite{Kuramoto75,Kuramoto84,Acebron}
\be \label{eq:dyn:Kuramoto}
\frac{d\theta_j}{dt} = \omega_j + \sum_{k=1}^N K_{jk} \sin(\theta_k-\theta_j),
\qquad j=1,\ldots, N
\ee
where we have defined the rescaled coupling constant
$K_{jk} \equiv \tilde{K}_{jk}/\gamma$.
Many applications of the Kuramoto model have been found, ranging from laser arrays or Josephson junctions, to neural networks and chemical oscillators
\cite{Acebron}.

\paragraph{Synchronization transition for globally coupled oscillators}
The mean-field version of the Kuramoto model consists in choosing uniform coupling constants $K_{jk}=K/N$ (the $1/N$ scaling is including to keep the interaction term bounded when $N$ increases), leading to the simplified equation
\be \label{eq-Kuram-oscMF}
\frac{d\theta_j}{dt} = \omega_j + \frac{K}{N}\sum_{k=1}^N \sin(\theta_k-\theta_j),
\qquad j=1,\ldots, N.
\ee
If the coupling is strong enough, the oscillators may enter a synchronized state, in which they all have the same frequency $\Omega$,
that is $d\theta_j/dt=\Omega$.
The frequency $\Omega$ can be evaluated using the property
\be
\sum_{j=1}^N \frac{d \theta_j}{dt} = \sum_{j=1}^N \omega_j \,,
\ee
from which the relation
\be
\Omega = \frac{1}{N} \sum_{j=1}^N \omega_j
\ee
follows. Hence if synchronization occurs, the common frequency of the oscillators is the arithmetic average of the natural frequencies $\omega_j$.
In view of introducing an order parameter for the synchronization transition, it is convenient to go in the ``rotating frame'' at frequency $\Omega$,
by defining $\theta_j'=\theta_j -\Omega  t$ and $\omega_j'=\omega_j-\Omega$.
In this way, Eq.~(\ref{eq-Kuram-oscMF}) is still valid for the primed variables, and $\sum_{j=1}^N \omega_j'=0$. In the following, we work with the variables $\theta_j'$ and $\omega_j'$, and drop the primes to lighten notations.

In these new variables, synchronization between oscillators is observed when the complex order parameter
\be \label{eq-Kuram-order-param}
r\,e^{i\psi} = \frac{1}{N} \sum_{j=1}^N e^{i\theta_j}
\ee
(or simply its modulus $r$) takes a non-zero value in the large $N$ limit.
In a stationary synchronized state, the phase $\psi$ takes a time-independent value, which is arbitrary since all phases $\theta_j$ can be shifted by an arbitrary amount, resulting in the same shift on the global phase $\psi$.
Using this complex order parameter, Eq.~(\ref{eq:dyn:Kuramoto}) can be rewritten as
\be \label{eq-Kuram-oscMF2}
\frac{d\theta_j}{dt} = \omega_j + Kr\sin(\psi-\theta_j),
\qquad j=1,\ldots, N,
\ee
where in general, $\psi$ depends on time.
In the large $N$ limit, on which we now focus, the set of natural frequencies $\omega_j$ is described by a density $g(\omega)$, chosen to be normalized according to $\int_{-\infty}^{\infty} g(\omega)d\omega=1$.
For the sake of simplicity, we assume in the following that $g(\omega)$
is even.
Although the dynamics of the oscillators is purely deterministic, it is possible to describe the dynamics of the phases $\theta_j$ by a probability distribution, if one starts from a set of initial conditions.
We denote as $\rho(\theta|\omega,t)$ the probability distribution of the phase $\theta$ of an oscillator of frequency $\omega$, starting from a set of initial conditions described by the distribution $\rho(\theta|\omega,0)$.
The distribution $\rho(\theta|\omega,t)$ evolves according to the following equation
\be \label{eq-Kuram-evol-rho}
\frac{\partial \rho}{\partial t}(\theta|\omega,t) + \frac{\partial}{\partial \theta}
\left[ \Big(\omega + Kr\sin(\psi-\theta)\Big) \rho(\theta|\omega,t)\right]=0.
\ee
Since we consider here the infinite $N$ limit, the order parameter $r e^{i\psi}$
is self-consistently determined as
\be \label{eq-Kuram-order-param2}
r\,e^{i\psi} = \langle e^{i\theta}\rangle \equiv
\int_{-\pi}^{\pi} d\theta \int_{-\infty}^{\infty} d\omega\,
e^{i\theta} \rho(\theta|\omega,t) g(\omega).
\ee
Looking for stationary solutions of the coupled equations (\ref{eq-Kuram-evol-rho}) and (\ref{eq-Kuram-order-param2}), one easily sees that the uniform distribution $\rho(\theta|\omega)=(2\pi)^{-1}$, which leads to $r=0$, is a solution for all values of the coupling constant $K$.
This solution obviously corresponds to the absence of synchronization.
Let us now look at possible synchronized solutions, which if they exist, are more likely to be present at high coupling $K$.
We thus assume that the order parameter $r e^{i\psi}$ is non-zero,
that is, $r>0$, and set $\psi=0$ without loss of generality.
If $|\omega| < Kr$, $\theta_0 = \sin^{-1}(\omega/Kr)$ is the stable fixed point of Eq.~(\ref{eq-Kuram-oscMF2}), resulting in the stationary distribution
\be
\rho(\theta|\omega) = \delta(\theta -\theta_0).
\ee
For $|\omega| > Kr$, no fixed point exists, and the dynamics takes the form
of `travelling' solutions.
The steady-state distribution $\rho(\theta|\omega)$ is obtained from Eq.~(\ref{eq-Kuram-evol-rho}) as
\be \label{eq-Kuram-rho-theta}
\rho(\theta|\omega) = \frac{1}{2\pi} \, \frac{\sqrt{\omega^2 -(Kr)^2}}{|\omega
-Kr\sin\theta|}.
\ee
The self-consistency equation (\ref{eq-Kuram-order-param2}) then reads
\be
r = \int_{-\pi}^{\pi} d\theta \int_{-Kr}^{Kr} d\omega\, e^{i\theta}
\delta\left(\theta-\sin^{-1}(\omega/Kr)\right)\, g(\omega) \,.
\label{eq-Kuram-self-consist}
\ee
Note that the contribution of Eq.~(\ref{eq-Kuram-rho-theta}) to the integral in Eq.~(\ref{eq-Kuram-self-consist}) vanishes by symmetry.
Using the change of variables $\omega=Kr\sin x$, it is easy to show (assuming that $g(\omega)$ is maximal at $\omega=0$) that Eq.~(\ref{eq-Kuram-self-consist}) has a solution $r>0$ when
\be
K > K_c \equiv \frac{2}{\pi g(0)} \,.
\ee
Hence a synchronization transition occurs at $K=K_c$, and the oscillators are synchronized for $K>K_c$. The synchronization effect can in particular be quantified by determining the scaling of the order parameter $r$ with $K-K_c$ close to the transition.
If $g(\omega)$ has a regular expansion to order $\omega^2$ around $\omega=0$,
one finds the generic scaling $r \sim (K-K_c)^{1/2}$ \cite{Acebron}.


\section{Local mean-field approximation}
\label{sec-MFFP}

The basic mean-field approximations that we have discussed up to now
focus on a single unit, and treat the influence of the rest of the system using self-consistent approximations. By doing so, spatial information is lost, since all units are considered as equivalent.
It is possible to improve such methods by retaining a spatial description, while still performing mean-field types of approximations which amount to neglecting correlations between local regions. We denote here such approaches under the generic term of local mean-field approximations.
Depending on the context, they may also be called mean-field Fokker-Planck or mean-field Smoluchowski equations, for instance.
In most cases, the local mean-field approximation consists in writing the equation for the $N$-body problem, and then assuming that the $N$-body distribution factorizes as a product of independent distributions associated with each degree of freedom. Note that these one-body distributions are not necessarily identical, which allows to retain some spatial dependence in the problem.

We provide below several explicit examples of the application of this approximation method. In Sect.~\ref{sec-KEP}, we come back to the description of the sheared elastoplastic model. Then we turn to systems of interacting self-propelled particles, either with short-range, velocity-aligning interactions in Sect.~\ref{sec-SPP-FP}, or with long-range hydrodynamic interactions in Sect.~\ref{sec-swimmers}. Although the detailed shape of the evolution equation differs from one case to another, it is generically found in these three examples that the evolution equation for the distribution $P$ of the local degree of freedom considered is quadratic in $P$, with either a local or non-local interaction kernel.

\subsection{Kinetic Elastoplastic Model}
\label{sec-KEP}

The so-called Kinetic Elastoplastic Model (KEP) \cite{Bocquet09} can be interpreted as a spatial version of the H\'ebraud-Lequeux model discussed
in Sect.~\ref{sec-HL}. 
Similarly to the H\'ebraud-Lequeux model \cite{HL98} and to its lattice generalizations \cite{Picard04,Picard05}, space is assumed to be divided into mesoscopic cells.
But while in the H\'ebraud-Lequeux model, the effect of the environment of a given mesoscopic cell was modeled phenomenologically as a diffusive term acting on the stress $\sigma$,
the idea of the KEP model is to retain a spatial description of the stress, and to compute explicitly the effect of other cells on any given cell.
As for the H\'ebraud-Lequeux model, a local plastic relaxation occurs with a probability $1/\tau$ per unit time when the local stress $\sigma_i$ exceeds the threshold $\sigma_c$. The local stress $\sigma_i$ then instantaneously drops to zero, before resuming an elastic phase. This plastic stress drop $\delta\sigma_i^{\rm pl} = -\sigma_i$ (where $\sigma_i$ is the value of the local stress just before the plastic event) is then elastically propagated to distant sites $j$ through the elastic propagator $G_{ji}$, leading to a stress variation $\delta \sigma_j = G_{ji}\delta\sigma_i^{\rm pl}$ on site $j$.

\subsubsection{Evolution equation for the local stress distribution}

In the framework of the local mean-field approximation,
the evolution of the distribution $P_i(\sigma,t)$ of the local stress $\sigma_i$ is given by
\bea \nonumber
\partial_t P_i(\sigma,t) &=& -\mu \dot{\gamma}_{\rm ext} \, \partial_{\sigma} P_i(\sigma,t)
- \frac{1}{\tau}\Theta(|\sigma|-\sigma_c)  P_i(\sigma,t) \\
&& \qquad \qquad \qquad \qquad \qquad
+ \Gamma_i(t)\, \delta(\sigma) + \mathcal{L}_i[{\bf P},{\bf P}](\sigma),
\label{eq:dist:Pisigma:KEP}
\eea
with ${\bf P} \equiv \{P_i\}$, and
where $\dot{\gamma}_{\rm ext}$ is the externally imposed shear rate, whereas $\Gamma_i(t)$ is the local plastic activity defined as
\be
\Gamma_i(t) = \frac{1}{\tau} \int_{|\sigma|>\sigma_c} P_i(\sigma,t) d\sigma.
\ee
The different terms in Eq.~(\ref{eq:dist:Pisigma:KEP}) have the same interpretation as in the H\'ebraud-Lequeux model discussed in Sect.~\ref{sec-HL}, except the last term $\mathcal{L}_i[{\bf P},{\bf P}](\sigma)$ which describes interactions between the focus cell $i$ and other cells.
In the H\'ebraud-Lequeux model, this term was phenomenologically replaced by a (self-consistent) diffusion term.
Here, interactions are modeled in a more detailed way.
Within the framework of the local mean-field hypothesis, one assumes that
the stresses $\sigma_i$ and $\sigma_j$ at sites $i \ne j$ are statistically independent. Then the interaction term can be written as a bilinear term,
\be \label{eq:KEP:LPP}
\mathcal{L}_i[{\bf P},{\bf P}](\sigma) = \frac{1}{\tau} \sum_{j (\ne i)} 
\int_{|\sigma'|>\sigma_c} d\sigma' P_j(\sigma',t) [P_i(\sigma- G_{ij}\sigma',t)
- P_i(\sigma,t)]
\ee
where $\delta \sigma_i \equiv - G_{ij}\sigma'$ is the stress variation on site $i$ generated by the plastic stress drop $\delta\sigma_j^{\rm pl} = -\sigma'$ due to a plastic event on site $j$.

The way to derive Eqs.~(\ref{eq:dist:Pisigma:KEP}) and (\ref{eq:KEP:LPP}) is to start from the master equation describing the evolution of the full $N$-body distribution $\mathcal{P}(\sigma_1,\dots,\sigma_N)$.
The local mean-field approximation then amounts to assuming that the distribution $\mathcal{P}(\sigma_1,\dots,\sigma_N)$ is factorized, namely
\be \label{eq:factor:KEP}
\mathcal{P}(\sigma_1,\dots,\sigma_N) = \prod_{i=1}^N P_i(\sigma_i),
\ee
with one-body distributions $P_i(\sigma_i)$ that depend on $i$ (without this $i$-dependence, the approximation would be a simple mean-field approximation rather than a local mean-field one).
Inserting the factorized form (\ref{eq:factor:KEP}) of $\mathcal{P}$ into the full master equation precisely leads to Eq.~(\ref{eq:dist:Pisigma:KEP}).

\subsubsection{Connection to the H\'ebraud-Lequeux model}

As a further approximation, it is convenient to assume that $\delta \sigma_i$
is small (which is correct as long as cells $i$ and $j$ are not too close),
and can be approximated as $\delta \sigma_i \approx G_{ij}\sigma_c$ (meaning that the stress before plastic relaxation is close to $\sigma_c$, which is true for small shear rates).
One can then expand $P_i(\sigma+\delta \sigma_i,t)$ to second order in $\delta \sigma_i$ as
\be
P_i(\sigma+\delta \sigma_i,t) \approx P_i(\sigma,t) + \delta \sigma_i \partial_{\sigma}P_i + \frac{1}{2} (\delta \sigma_i)^2 \partial_{\sigma}^2 P_i \,.
\ee
Eq.~(\ref{eq:dist:Pisigma:KEP}) can be rewritten as
\bea \nonumber
\partial_t P_i(\sigma,t) &=& -\mu \dot{\gamma}_i(t) \, \partial_{\sigma} P_i(\sigma,t)
- \frac{1}{\tau}\Theta(|\sigma|-\sigma_c)  P_i(\sigma,t) \\
&& \qquad \qquad \qquad \qquad \qquad
+ \Gamma_i(t)\, \delta(\sigma) + D_i(t) \partial_{\sigma}^2 P_i(\sigma,t)
\label{eq:dist:Pisigma:KEP2}
\eea
where
\bea
\dot{\gamma}_i(t) &=& \dot{\gamma}_{\rm ext} + \sigma_c \sum_{j (\ne i)} G_{ij} \Gamma_j(t)
\label{eq:gammadoti}\\
D_i(t) &=& \frac{1}{2} \sigma_c^2 \sum_{j (\ne i)} G_{ij}^2 \Gamma_j(t).
\label{eq:Di}
\eea
In the case where the system is spatially homogeneous (so that $D_i$ and $\Gamma_i$ are independent of $i$), one recovers the diffusion term of the stress introduced in the H\'ebraud-Lequeux model, with a diffusion coefficient proportional to the plastic activity, $D=\alpha \Gamma$. One advantage of this approach as compared to the original H\'ebraud-Lequeux model is that the proportionality coefficient $\alpha$ now has an explicit expression given by
Eq.~(\ref{eq:Di}), namely
\be
\alpha = \frac{1}{2} \sigma_c^2 \sum_{j (\ne i)} G_{ij}^2,
\ee
while this coefficient was purely phenomenological in the H\'ebraud-Lequeux model.
In addition, distant plastic events also induce a correction
to the shear rate, which thus takes the form of an effective local shear rate
$\dot{\gamma}_i(t)$ as given in Eq.~(\ref{eq:gammadoti}).
Note that the correction vanishes for a homogeneous system because
$\sum_{j (\ne i)} G_{ij}=0$ (a property of the elastic propagator),
thus recovering the H\'ebraud-Lequeux model. Corrections to the shear rate are however present when the system is inhomogeneous, for instance during transient states, or in presence of shear banding.

The properties of the system can then be deduced from Eq.~(\ref{eq:dist:Pisigma:KEP2}) using essentially the same methods as for the H\'ebraud-Lequeux model, and taking into account the non-local self-consistency relations Eqs.~(\ref{eq:gammadoti}) and 
(\ref{eq:Di}). This can be more conveniently done by taking first a continuous limit, in which in particular the diffusion coefficient $D_i(t)$ becomes
a function $D({\bf r},t)$ of the continuous space variable ${\bf r}$,
satisfying
\be \label{eq:stress:diff:KEP}
D({\bf r},t) = m \Delta \Gamma({\bf r},t) + \alpha \Gamma({\bf r},t)
\ee
where $\Delta$ is the Laplacian operator, and $m$ is a parameter that can be expressed in terms of the elastic propagator \cite{Bocquet09}.
Again, it appears clearly that for a spatially homogeneous system,
the results of the H\'ebraud-Lequeux model are recovered.
However, the present KEP model allows one to account for spatial heterogeneities, which play an important role in confined geometries like microchannels \cite{Bocquet09}.
The predicted non-local effects resulting from the presence of a Laplacian term in Eq.~(\ref{eq:stress:diff:KEP}) have been confirmed in experiments \cite{Colin10} as well as in numerical simulations of an elastoplastic lattice model
\cite{Nicolas13}.

\subsection{Self-propelled particles with short-range aligning interactions}
\label{sec-SPP-FP}

As already mentioned above, self-propelled particles are a simple model aiming at describing physical systems like active colloids as well as, at a more qualitative level, biological systems like colonies of myxobacteria, flocks of birds or schools of fish \cite{Marchetti-RMP}.
In all cases, the particle carries a heading vector (an orientation) and is subjected to a propelling force, often of constant amplitude, acting along this heading vector.
If the relaxation of the velocity is fast with respect to other time scales of the dynamics (like the time to travel the typical distance between particles), one may consider that the velocity is along the heading vector, and that 
the speed takes a constant value $v_0$.

A situation of interest is when self-propelled particles have interactions that tend to align their velocity vectors.
This is exemplified through the well-known Vicsek model \cite{Vicsek95,Chate04}, in which a particle takes at the next time step the average direction, up to some noise, of all particles (including itself) situated within the interaction range.
This rule results in a competition between alignment and noise, so that collective motion (i.e., polar alignment) sets in below a density-dependent noise threshold. At higher noise, the system remains isotropic, and no collective motion is observed.

\begin{figure}[t!]
\centering\includegraphics[width=7.5cm]{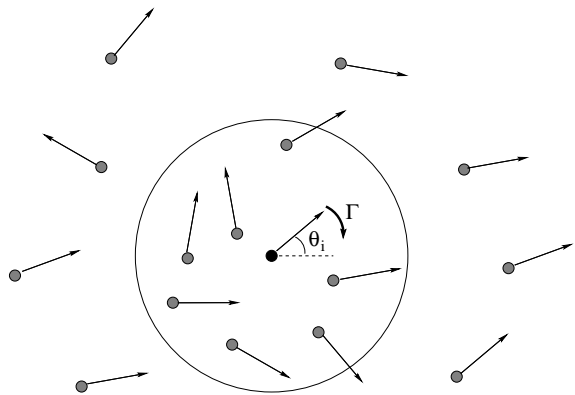}
\caption{Self-propelled particles with finite-range velocity alignment interactions. The interaction range is indicated by a circle around the focus particles $i$ (in black).
Interactions with other particles within the circle generate a torque $\Gamma$, which changes the direction $\theta_i$ of the velocity of particle $i$ to align it on the average direction of the velocities of neighboring particles.}
\label{fig-SPP-FP}
\end{figure}

It is tempting to describe such a system through a local mean-field approach \cite{Farrell12,Peruani,Grossmann13,Grossmann14}.
Note that similar approaches have also been used for self-propelled particles with nematic interactions \cite{Marchetti08a,Marchetti08b,Peruani,Bertin15}.
For concreteness, let us consider a simple two-dimensional model of $N$ aligning self-propelled particles of position ${\bf r_i}$ moving at constant speed $v_0$. The velocity vector is then simply defined by its angle $\theta_i$ with
respect to a reference direction.
The model is governed by the following overdamped dynamics,
\bea \label{eq:SPP:Langevin:r}
\frac{d {\bf r}_i}{dt} &=& v_0 {\bf e}(\theta_i) + {\boldsymbol \xi}_i(t)\\
\frac{d \theta_i}{dt} &=& \sum_{j=1}^N \Gamma(\theta_j-\theta_i, {\bf r}_j -{\bf r}_i) + \eta_i(t)
\label{eq:SPP:Langevin:theta}
\eea
where ${\bf e}(\theta)$ is the unit vector of direction $\theta$,
and $\Gamma(\Delta\theta,\Delta {\bf r})$ is the alignment torque with the neighboring particles ---the friction coefficient has been set to one.
$\Gamma(\Delta\theta,\Delta {\bf r})$ is assumed to be $2\pi$-periodic with respect to $\Delta\theta$, and to go to zero when $||\Delta {\bf r}||$ goes to infinity.
A simple explicit expression for $\Gamma(\Delta\theta,\Delta {\bf r})$ can be
\be \label{eq:def:torque}
\Gamma(\Delta\theta,\Delta {\bf r}) = \frac{\sin \Delta\theta}{\pi R_0^2} 
\, \Theta(R_0-||\Delta {\bf r}||)
\ee
where $\Theta$ is the Heaviside function, and $R_0$ is the interaction range.
Finally, ${\boldsymbol \xi}(t)$ and $\eta(t)$ are positional and angular white noises, with respective diffusion coefficients $D$ and $D_R$:
\bea
\la \xi_{i,\alpha}(t) \ra &=&0, \qquad \la \xi_{i,\alpha}(t)\xi_{j,\beta}(t') \ra = 2D \delta_{ij} \delta_{\alpha\beta} \delta(t-t'),\\
\la \eta_i(t) \ra &=& 0, \qquad
\la \eta_i(t)\eta_j(t') \ra = 2D_R\,\delta_{ij} \delta(t-t').
\eea
where $\alpha,\beta=1,2$ denote Cartesian coordinates.
A sketch of the model is given in Fig.~\ref{fig-SPP-FP}.

\subsubsection{Mean-field equation for the one-particle phase space distribution}

Let us provide in this case the explicit derivation of the mean-field equation for the one-particle phase space distribution.
This type of approach is generic and can be applied to many different cases.
We start by writing the exact evolution equation for the $N$-body distribution $P_N({\bf r}_1,\theta_1,\dots,{\bf r}_N,\theta_N,t)$.
This evolution equation is nothing but the Fokker-Planck equation associated with the Langevin dynamics
given by Eqs.~(\ref{eq:SPP:Langevin:r}) and (\ref{eq:SPP:Langevin:theta}),
namely
\bea \nonumber
&& \frac{\partial P_N}{\partial t} 
+ \sum_{i=1}^N \frac{\partial}{\partial {\bf r}_i} \cdot [v_0 {\bf e}(\theta_i) P_N] + \sum_{i=1}^N \frac{\partial}{\partial \theta_i} \left(
\sum_{j (\ne i)} \Gamma(\theta_j-\theta_i,{\bf r}_j-{\bf r}_i) P_N \right) \\
&& \qquad \qquad \qquad \qquad \qquad  = \sum_{i=1}^N D \left( \frac{\partial}{\partial {\bf r}_i} \right)^2 P_N
+ \sum_{i=1}^N D_R \frac{\partial^2 P_N}{\partial \theta_i^2} \,.
\label{eq:Nbody:SPP:LMF}
\eea
A systematic way to perform the mean-field approximation is to assume that
the $N$-body distribution factorizes as
\be
P_N({\bf r}_1,\theta_1,\dots,{\bf r}_N,\theta_N,t) = \prod_{i=1}^N P_1({\bf r}_i,\theta_i,t) \,,
\ee
where $P_1$ is normalized such that $\int d{\bf r} d\theta P_1({\bf r},\theta,t) =1$.
Integrating Eq.~(\ref{eq:Nbody:SPP:LMF}) on the variables 
$({\bf r}_2,\theta_2,\dots,{\bf r}_N,\theta_N)$, we end up with\footnote{We have assumed that the space integral of all divergence terms is equal to zero. This can be justified either by assuming periodic boundary conditions in space, or by assuming that $P_1$ and its derivatives with respect to ${\bf r}$ vanish when $||{\bf r}||$ goes to infinity.}
\bea \nonumber
&& \frac{\partial P_1}{\partial t}({\bf r}_1,\theta_1,t) 
+ \frac{\partial}{\partial {\bf r}_1} \cdot [v_0 {\bf e}(\theta_1) P_1({\bf r}_1,\theta_1,t)] \\ \nonumber
&& \qquad \quad + \frac{\partial}{\partial \theta_1} \left(
\sum_{j=2}^N \int d{\bf r}_j d\theta_j \,
\Gamma(\theta_j-\theta_1,{\bf r}_j-{\bf r}_1) P_j({\bf r}_j,\theta_j,t) P_1({\bf r}_1,\theta_1,t) \right) \\
&& \qquad \quad = D \left( \frac{\partial}{\partial {\bf r}_1} \right)^2 P_1({\bf r}_1,\theta_1,t) + D_R \frac{\partial^2 P_1}{\partial \theta_1^2}({\bf r}_1,\theta_1,t) \,.
\label{eq:Nbody:SPP:LMF2}
\eea
The integral in the lhs of Eq.~(\ref{eq:Nbody:SPP:LMF2}) is independent of $j$, so that the sum simply yields a factor $N-1$ in front of the integral.

In practice, it is then convenient to work with the one-body phase-space distribution $f({\bf r},\theta,t)=N P_1({\bf r},\theta,t)$, so that
$\int d\theta\, f({\bf r},\theta,t)$ is the local density $\rho({\bf r},t)$ of particles.
With this change of normalization, Eq.~(\ref{eq:Nbody:SPP:LMF2}) can be rewritten in the limit $N \to \infty$ in the form \cite{Farrell12}
\be
\frac{\partial f}{\partial t} + \nabla \cdot [v_0 {\bf e}(\theta) f]
+ \frac{\partial}{\partial \theta} [ \overline{\Gamma}_{[f]} f]
= D \Delta f + D_R \frac{\partial^2 f}{\partial \theta^2},
\label{eq:MFFP}
\ee
where the local average torque $\overline{\Gamma}_{[f]}({\bf r},t)$ exerted on the focus particle by particles located within the interaction range is given by
\be
\overline{\Gamma}_{[f]}({\bf r},t) = 
\int_{-\pi}^{\pi} d\theta' \int d{\bf r}' f({\bf r}',\theta')\,
\Gamma(\theta'-\theta,{\bf r}'-{\bf r}) \,.
\ee
This average torque  is computed by weighting the individual torque $\Gamma(\theta'-\theta,{\bf r}'-{\bf r})$ by the local phase-space density $f({\bf r}',\theta')$. This approximation is thus supposed to be relevant in the case where many particles are present within the interaction range.
The notation $\overline{\Gamma}_{[f]}$ emphasizes the functional dependence of the average torque on the one-particle density $f$, making Eq.~(\ref{eq:MFFP}) a non-linear (quadratic) equation in $f$.
Note also that the number $N$ of particles no longer explicitly appears
in Eq.~(\ref{eq:MFFP}).

One can check that the stationary uniform angular distribution
$f(\mathbf{r},\theta)=\rho/2\pi$ is a solution of
Eq.~(\ref{eq:MFFP}) for an arbitrary constant density $\rho$.
To analyze the behaviour of Eq.~(\ref{eq:MFFP}) beyond this simple uniform solution, it is convenient to expand $f({\bf r},\theta,t)$ into angular Fourier modes,
\bea
f({\bf r},\theta,t) &=& \frac{1}{2\pi} \sum_{k=-\infty}^{\infty}
\hat{f}_k({\bf r},t) \, e^{-ik\theta},\\
\hat{f}_k({\bf r},t) &=& \int_{-\pi}^{\pi} f({\bf r},\theta,t)
\, e^{ik\theta} d\theta.
\eea
Eq.~(\ref{eq:MFFP}) then transforms into
\be \label{eq:MFFP:Fourier}
\frac{\partial \hat{f}_k}{\partial t}
+ \frac{v_0}{2} (\hat{\triangledown} \hat{f}_{k-1} + \hat{\triangledown}^* \hat{f}_{k+1})
= -D_R k^2 \hat{f}_k + \frac{i\gamma k}{2\pi} \sum_{m=-\infty}^{\infty}
\hat{\Gamma}_{-m} \hat{f}_m \hat{f}_{k-m}
\ee
where $\hat{\triangledown}$ and $\hat{\triangledown}^*$ are the complex differential operators
\be \label{eq:def:nabla:complex}
\hat{\triangledown} = \frac{\partial}{\partial x} + i \frac{\partial}{\partial y},
\qquad 
\hat{\triangledown}^* = \frac{\partial}{\partial x} - i \frac{\partial}{\partial y},
\ee
and where $\hat{\Gamma}_k$ is the angular Fourier coefficient of $\Gamma(\Delta\theta,\Delta {\bf r})$:
\be
\hat{\Gamma}_k({\bf r}) = \int_{-\pi}^{\pi} \Gamma(\theta,{\bf r}) \, e^{ik\theta} d\theta.
\ee
Note that in Eq.~(\ref{eq:MFFP:Fourier}), $\hat{f}_k({\bf r'})$
has been approximated by $\hat{f}_k({\bf r})$, assuming that $f({\bf r},\theta,t)$
has only tiny variations over the interaction range.
Corrections to this approximation can be systematically derived by expanding
$\hat{f}_k({\bf r'})$ in powers of ${\bf r'}-{\bf r}$.

Note that the method used here is quite general, and can be used for instance for different types of self-propelled particles.
It has also been used for example for self-propelled particles interacting via pairwise forces deriving from a potential, but without aligning interactions
\cite{Speck13,Speck15}.

\subsubsection{Derivation of hydrodynamic equations}

The field $\hat{f}_0({\bf r})$ can be identified with the density field $\rho({\bf r})$, so that Eq.~(\ref{eq:MFFP:Fourier}) yields for $k=0$
\be \label{eq:contin:MFFP:compl}
\frac{\partial \rho}{\partial t} + v_0 \, {\rm Re}(\hat{\triangledown}^* \hat{f}_1) =0
\ee
with ${\rm Re}(z)$ the real part of the complex number $z$.
Going back to vectorial notations, Eq.~(\ref{eq:contin:MFFP:compl})
is nothing but the usual continuity equation
\be \label{eq:contin:SPP}
\frac{\partial \rho}{\partial t} + \nabla \cdot (\rho {\bf v}) =0
\ee
where the hydrodynamic velocity field is defined as
\be
{\bf v}({\bf r},t) = \frac{v_0}{\rho({\bf r},t)}
\int_{-\pi}^{\pi} d\theta \, f({\bf r},\theta,t) \, {\bf e}(\theta).
\ee
The evolution equation for the velocity field can be obtained from Eq.~(\ref{eq:MFFP:Fourier}) for $k=1$. At linear order in $\hat{f}_1$, one finds an instability towards collective motion (nonzero values of $\hat{f}_1$) below a density-dependent threshold value of the noise. It is necessary to include non-linear terms to saturate the instability. This is done minimally by taking into account the equation for $\hat{f}_2$ and assuming that $\hat{f}_2$ is slaved to $\hat{f}_1$ while higher order angular modes like $\hat{f}_3$ can be neglected
(note that in the case where $\Gamma$ is given by Eq.~(\ref{eq:def:torque}), no higher order term appears). A more precise justification of this procedure is obtained through the scaling assumption
\cite{Farrell12,EPJST14,Bertin15}
\be \label{eq:scaling:SPP}
\hat{f}_k \sim \epsilon^{|k|} , \qquad  \partial_t \sim \partial_x \sim \partial_y \sim \epsilon \,,
\ee
where $\epsilon$ is a small parameter encoding the distance to the instability threshold.
Eliminating $\hat{f}_2$ in the equation for $\hat{f}_1$, we end up with a closed equation for the fields $\hat{f}_1$ and $\rho$.
Mapping complex numbers onto two-dimensional vectors, one eventually obtains the following hydrodynamic equations for the ``momentum'' field\footnote{Note that momentum is not conserved in this model, unlike in standard fluids. Hence the reason for keeping momentum in the hydrodynamic description is that it is the variable associated with the spontaneous breaking of the rotational symmetry.}
${\bf w}=\rho {\bf v}$,
\bea \label{eq-w}
\frac{\partial \mathbf{w}}{\partial t} + \gamma (\mathbf{w} \cdot \nabla)
\mathbf{w} &=& -\frac{1}{2}\nabla (\rho - \kappa \mathbf{w}^2) \\
\nonumber
&+& (\mu - \xi \mathbf{w}^2) \mathbf{w}
+ \nu \Delta \mathbf{w} - \kappa (\nabla \cdot \mathbf{w}) \mathbf{w}.
\eea
where $\Delta$ is the Laplacian operator, and where
all coefficients $\gamma$, $\kappa$, $\mu$, $\xi$ and $\nu$ are known as a function of microscopic parameters of the model and of the density \cite{Farrell12} ---see also \cite{EPJST14} and \cite{Bertin15} for related problems.
Note that the coefficient $\mu$ changes sign as a function of both density and noise amplitude $D_R$. It is negative at low density or high noise, and positive at higher density and low noise.
When $\mu$ is positive, the uniform motionless solution ${\bf w} =0$ becomes unstable, and a solution with uniform collective motion $||{\bf w}||=\sqrt{\mu/\xi}$ emerges. This solution, however, is itself unstable close to the transition to collective motion \cite{EPJST14}, leading to the generic emergence of solitary waves \cite{CaussinPRL15,CaussinPRE15,BDG09}.

\subsection{Microswimmers with long-range hydrodynamic interactions}
\label{sec-swimmers}

The systems of interacting self-propelled particles described in Sect.~\ref{sec-SPP-FP} were 'dry' systems, in which the effect of the fluid surrounding the particles is neglected.
This is often the case with two-dimensional systems, where particles are in contact with a solid substrate \cite{Marchetti-RMP}.
In three dimensional systems, the situation is different, and the fluid plays an essential role as no solid substrate is present.
Self-propelled particles are called ``swimmers'' in this context, and they experience long-range interactions between them that are mediated by the fluid.
This is due to the fact that particles are advected by the fluid flow, and that a given swimmer, being a force dipole, creates a long-range disturbance of the fluid flow around itself.

\subsubsection{Swimming in the flow generated by the other particles}

We assume that a swimmer moves at constant speed $v_0$ along its heading vector ${\bf n}$. Its velocity in the rest frame is thus, taking into account the advection by the fluid flow ${\bf u}({\bf r},t)$,
\be \label{eq:drdt:swim}
\frac{d {\bf r}_i}{dt} = v_0 {\bf n}_i + {\bf u}({\bf r}_i,t) + {\boldsymbol \xi}_i(t)
\ee
where ${\boldsymbol \xi}(t)$ is a white noise with diffusion coefficient $D$,
\be
\la \xi_{i,\alpha}(t) \ra =0, \qquad \la \xi_{i,\alpha}(t)\xi_{j,\beta}(t') \ra = 2D \delta_{ij} \delta_{\alpha\beta} \delta(t-t')
\ee
($\alpha,\beta$ denote Cartesian coordinates).
Modeling the torque exerted by the flow on the swimmers through Jeffery's model \cite{Jeffery}, the time derivative of the heading vector ${\bf n}$
is given by
\be \label{eq:dndt:swim}
\frac{d {\bf n}_i}{dt} = \Gamma({\bf n}_i,{\bf r}_i,t)
\equiv ({\bf I}-{\bf n}_i{\bf n}_i) \cdot \big(\gamma {\bf E}({\bf r}_i,t) + {\bf W}({\bf r}_i,t) \big) \cdot {\bf n}_i ,
\ee
where ${\bf I}$ is the unit tensor, ${\bf E}$ and ${\bf W}$ are respectively the rate-of-strain and vorticity tensors,
\be \label{eq:swim:EW}
{\bf E} = \frac{1}{2}(\nabla {\bf u} + \nabla {\bf u}^{\bf T}), \qquad
{\bf W} = \frac{1}{2}(\nabla {\bf u} - \nabla {\bf u}^{\bf T}),
\ee
and $\gamma$ is a shape parameter ($\gamma \approx 1$ for rods) \cite{Shelley08a}. Note that angular noise could also be taken into account, but we neglect it here for simplicity.

Integrating the dynamics of the swimmers, as defined by Eqs.~(\ref{eq:drdt:swim}) and (\ref{eq:dndt:swim}), requires the knowledge of the hydrodynamic velocity
field ${\bf u}({\bf r},t)$, from which the rate-of-strain and vorticity tensors can also be deduced.
We assume that the velocity field is generated by the swimmers themselves, and that boundary conditions impose no flow.
Swimmers are typically micrometric objects swimming in water, so that the Reynolds number is very low. In this situation,
the velocity field ${\bf u}$ is governed by the Stokes equation
\be \label{eq:Stokes}
-\nabla p + \eta \Delta {\bf u} + \nabla \cdot {\boldsymbol \sigma}_a = 0
\ee
where $p$ is the pressure, $\eta$ is the dynamic viscosity of the fluid, and
${\boldsymbol \sigma}_a$ is the active stress tensor generated by the assembly of swimmers
\be \label{eq:def:sigma-a}
{\boldsymbol \sigma}_a ({\bf r},t) = \sum_i \sigma_0 \left({\bf n}_i{\bf n}_i
-\frac{1}{3}{\bf I}\right) \delta({\bf r}_i -{\bf r})
\ee
with $\sigma_0$ the magnitude of the force dipole of each swimmer.
Note that the sign of $\sigma_0$ is an important characteristics;
Swimmers with $\sigma_0>0$ are called pullers, while swimmers with $\sigma_0<0$ are called pushers. The collective behaviour of pushers and pullers is often quite different \cite{Shelley08a,review-swim}. For instance, the stability properties of the isotropic state differs in both cases, as discussed below.
Note also that from Eqs.~(\ref{eq:Stokes}) and (\ref{eq:def:sigma-a}),
the velocity field ${\bf u}({\bf r},t)$ is implicitly a function of the positions of all particles; The same is also true for the torque $\Gamma({\bf n}_i,{\bf r}_i,t)$.

\subsubsection{Statistical description in the local mean-field approximation}

The statistical description of interacting swimmers in three-dimensions has been addressed in a series of works \cite{Shelley08a,Shelley08b,Marchetti09,Shelley10,Shelley13} ---see also \cite{Brotto13} for the quasi-two-dimensional case when particles are strongly confined in the third direction.
In order to describe statistically a large assembly of $N$ swimmers, one can introduce the one-particle distribution $f({\bf r}, {\bf n}, t)$ as done in the previous examples of this section ---note that one keeps here the three-dimensional heading vector as it cannot be described by a single angle as in two dimensions.

Similarly to the derivation made in Sect.~\ref{sec-SPP-FP}, 
the local mean-field approximation consists in starting
from the evolution equation of the $N$-body distribution $P_N$,
assuming that it is factorized as $P_N({\bf r}_1,\theta_1,\dots,{\bf r}_N,\theta_N)=\prod_i [f({\bf r}_i,\theta_i)/N]$.
One then integrates the resulting equation over the variables
$({\bf r}_2,\theta_2,\dots,{\bf r}_N,\theta_N)$, leading to the following
equation for $f({\bf r}, {\bf n}, t)$,
\be \label{eq:f:swim}
\frac{\partial f}{\partial t} + \nabla \cdot \left[ \big(v_0{\bf n}+ \overline{\bf u}({\bf r},t) \big) f\right] + \nabla_{\bf n} \cdot \big(\overline{\Gamma}({\bf r},t) f) = D \Delta f \,.
\ee
The notation $\nabla_{\bf n}$ indicates the gradient with respect to the vector ${\bf n}$ of fixed norm $||{\bf n}||=1$ (gradient on the unit sphere).
The velocity field $\overline{\bf u}({\bf r},t)$ is the ensemble average, over the positions of all swimmers, of the velocity field generated by the swimmers.
In a similar way, $\overline{\Gamma}({\bf r},t)$ denotes the average torque resulting from the average over the positions of the swimmers.

Thanks to the linearity of the Stokes equation (\ref{eq:Stokes}), the average velocity field $\overline{\bf u}({\bf r},t)$ is obtained by solving the average Stokes equation
\be \label{eq:Stokes:av}
-\nabla \overline{p} + \eta \Delta \overline{\bf u} + \nabla \cdot \overline{\boldsymbol \sigma}_a = 0
\ee
describing the flow generated by the average active stress,
\be
\overline{\boldsymbol \sigma}_a = \int \sigma_0 \, \left({\bf n}{\bf n} -\frac{1}{3}{\bf I}\right) f({\bf r},{\bf n},t)\, d{\bf n} \,.
\ee
The quantity $\overline{p}$ in Eq.~(\ref{eq:Stokes:av}) is the average pressure field, that does not need to be determined explicitly, but ensures the incompressibility of the flow.
The solution of the average Stokes equation is then a linear functional
$\overline{\bf u}_{[f]}({\bf r},t)$ of the one-particle probability distribution
$f({\bf r},{\bf n},t)$, due to the linearity of the Stokes equation
and to the linear dependence of $\overline{\boldsymbol \sigma}_a$ on $f$.
The average torque $\overline{\Gamma}({\bf r},t)$, obtained by averaging
Eqs.~(\ref{eq:dndt:swim}) and (\ref{eq:swim:EW}) over the positions of the swimmers, also becomes a linear functional $\overline{\Gamma}_{[f]}({\bf r},t)$ of the distribution $f$.
The evolution equation for $f$, Eq.~(\ref{eq:f:swim}), then becomes a quadratic equation in $f$,
\be \label{eq:f:swim:MF}
\frac{\partial f}{\partial t} + \nabla \cdot [(v_0{\bf n}+\overline{\bf u}_{[f]}) f] +
\nabla_{\bf n} \cdot (\overline{\Gamma}_{[f]} f) = D \Delta f \,.
\ee
The mathematical form of Eq.~(\ref{eq:f:swim:MF}) is thus qualitatively
similar to Eq.~(\ref{eq:MFFP}) studied in the case of 'dry' self-propelled particles discussed in Sect.~\ref{sec-SPP-FP}.
Yet, at variance with this previous case, Eq.~(\ref{eq:f:swim:MF}) is now fully non-local due to the non-locality of ${\bf u}_{[f]}$ and $\Gamma_{[f]}$.
In contrast, Eq.~(\ref{eq:MFFP}) is only weakly non-local as it only involves the variation of $f$ over the interaction range, which may be evaluated using low order space derivatives of $f$.
The equation was even turned into a local equation by neglecting variations of $f$ over the interaction range ---see Eq.~(\ref{eq:MFFP:Fourier}).

Stability analyses have been performed in different limits using Eq.~(\ref{eq:f:swim:MF}). It has been shown in particular that in the absence of diffusion ($D=0$), fully aligned suspensions are unstable stationary states of the dynamics
\cite{Shelley08a,Simha02}.
The stability of isotropic suspensions has also been studied,
leading to an instability of the isotropic state for suspensions of pushers,
while the isotropic state for suspensions of pullers is stable \cite{Shelley08a}.


\section{Kinetic theory and Boltzmann equation}
\label{sec-kinetic-th}

In the previous section, we have discussed how the spatial dependence can be included in a mean-field description. Generally speaking, the idea is to perform a local average over the interaction range, which leads to a non-linear (quadratic) equation for the one-particle phase space distribution.
This local mean-field approximation is supposed to be valid when particles have persistent interactions (as opposed to instantaneous interactions like collisions) over an interaction range that is larger than the typical interparticle distance, so that a significant number of particles are interacting with the focus particle at any time, thus justifying the evaluation of a local average force.
Long range-interactions, like hydrodynamic interactions between swimmers, are particularly well-suited for this type of approximation, leading in this case to non-local equations.

In the opposite limit where the interaction range is small with respect to the interparticle distance (an extreme case being that of hard spheres or discs), interactions occur through collisions that are very localized in space and time, and it is not justified to replace these collisions by a continuous average force. Instead, one has to take into account the probability of collision per unit time, and to make a statistical balance of the changes of physical quantities (velocities,...) during collisions, which are in general assumed to be restricted to binary collisions.
This is the purpose of kinetic theory, which we will consider below only in the restricted form of the Boltzmann equation \cite{Poschel04,Puglisi-book}. More involved treatments can be found for instance in \cite{Poschel04}.

We discuss here two paradigmatic examples, namely the kinetic theory of a driven granular gas (Sect.~\ref{sec-kinth-grangas}) and that of self-propelled particles with velocity alignment interactions (Sect.~\ref{sec-SPP-Boltz}).
From a formal point of view, the Boltzmann equation governing the one-body phase-space distribution contains a quadratic non-linearity with a local kernel, and as such shares some similarities with the equations obtained from the mean-field local approximation (Sect.~\ref{sec-MFFP}). This formal analogy will appear clearly in the case of interacting self-propelled particles, where results of both approaches can be compared (see also \cite{Bertin15}).

\subsection{Driven granular gas}

\label{sec-kinth-grangas}

The example of the granular gas has already been discussed in a simplified mean-field framework in Sect.~\ref{sec-granul-gas}.
Here, we discuss the statistics of the granular gas in the more refined framework of kinetic theory and Boltzmann equation.
A granular gas is a gas of inelastic particles that evolves through binary collisions \cite{Haff,Sela95,Shapiro95,Brey97,Trizac99,Trizac02,Trizac06,Puglisi98,Gradenigo11}. 
Interestingly, the convergence to a stationary state under the combined
effect of injection and dissipation of energy does not satisfy the usual
H-theorem \cite{Droz06}, but can be described by
a generalized H-theorem \cite{Puglisi13}.

\begin{figure}[t!]
\centering\includegraphics[height=4cm]{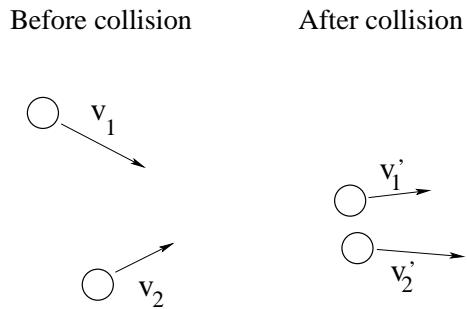}
\caption{Illustration of the binary collision of inelastic grains.
The total momentum is conserved during the collision, but not the kinetic energy.}
\label{fig-kinth-gran}
\end{figure}

In the following, we assume that all particles are identical. Each collision conserves the total momentum of the colliding particles, but dissipates a fraction of their kinetic energy (see Fig.~\ref{fig-kinth-gran}).
Neglecting the rotational degrees of freedom,
dissipative collisions are implemented
for hard disks or hard spheres by assuming the components of the velocities along the line joining the centers of the two colliding particles to obey the inelastic reflection law
${\bf v}_{12}^* \cdot \hat{\boldsymbol \sigma}
= -\alpha {\bf v}_{12} \cdot \hat{\boldsymbol \sigma}$,
with ${\bf v}_{12} = {\bf v}_1 - {\bf v}_2$, and where the star denotes postcollisional quantities; $\hat{\boldsymbol \sigma}$ is a unit vector along the direction joining the centers of the two particles, and collisions occur under the condition ${\bf v}_{12} \cdot \hat{\boldsymbol \sigma}>0$.
The coefficient $0<\alpha<1$ is called the (normal) restitution factor.
The other component, which is along the total momentum of the two particles, is unchanged thus ensuring momentum conservation.
The post-collisional velocities ${\bf v}_1^*$ and ${\bf v}_2^*$
are thus given by \cite{Ernst98}
\bea
{\bf v}_1^* &=& {\bf v}_1 - \frac{1}{2} (1+\alpha) ({\bf v}_{12} \cdot \hat{\boldsymbol \sigma}) \hat{\boldsymbol \sigma}\\
{\bf v}_2^* &=& {\bf v}_1 + \frac{1}{2} (1+\alpha) ({\bf v}_{12} \cdot \hat{\boldsymbol \sigma}) \hat{\boldsymbol \sigma}.
\eea

\subsubsection{Boltzmann equation for the driven granular gas}

In the absence of external forces acting on the particles, the evolution of the one-particle phase space distribution $f({\bf r},{\bf v},t)$ is described by the Boltzmann equation\footnote{A deterministic external force, like gravity, may also be included by adding to the l.h.s.~of Eq.~(\ref{eq:Boltz:gran})
a term $\nabla_{\bf v} \cdot ({\bf a} f)$, where ${\bf a}$ is the accerelation generated by the external force (that is, the force divided by the particle mass) and $\nabla_{\bf v}$ is the gradient with respect to velocity components.} \cite{Ernst98}
\be \label{eq:Boltz:gran}
\frac{\partial f}{\partial t} + \nabla \cdot ({\bf v} f) = I[f,f]
\ee
where $I[f,f]$ is a bilinear functional of $f$ describing the collision process,
\be
I[f,f] = r_0^{d-1} \int d{\bf v}_2 \int d\hat{\boldsymbol \sigma}\, \Theta({\bf v}_{12} \cdot \hat{\boldsymbol \sigma}) \left[ \frac{1}{\alpha^2}
f({\bf v}_1^{**},t) f({\bf v}_2^{**},t) - f({\bf v}_1,t) f({\bf v}_2,t) \right]
\ee
with $r_0$ the diameter of the particles, and where ${\bf v}_i^{**}$ are the precollisional velocities associated with given postcollisional velocities ${\bf v}_i$,
\bea
{\bf v}_1^{**} &=& {\bf v}_1 - \frac{1}{2} \left(1+\frac{1}{\alpha}\right) ({\bf v}_{12} \cdot \hat{\boldsymbol \sigma}) \hat{\boldsymbol \sigma}\\
{\bf v}_2^{**} &=& {\bf v}_1 + \frac{1}{2} \left(1+\frac{1}{\alpha}\right) ({\bf v}_{12} \cdot \hat{\boldsymbol \sigma}) \hat{\boldsymbol \sigma}.
\eea
The Boltzmann equation is derived using the approximation that colliding particles are not correlated before collision.
This approximation is supposed to be valid in the low density limit.
For moderate density, it is possible to improve the quality of the approximation by taking into account correlations on the positions that are induced by steric effects. This is done by including a factor $\chi(\rho)$ in the collision integral $I[f,f]$, where $\chi(\rho)$ is the pair correlation function of hard spheres or disks at contact as a function of the density
\be \label{eq:density:gran}
\rho({\bf r},t)=\int d{\bf v} f({\bf r},{\bf v},t).
\ee
The corresponding generalization of the Boltzmann equation,
\be \label{eq:Enskog:Boltz:gran}
\frac{\partial f}{\partial t} + \nabla \cdot ({\bf v} f) = \chi(\rho) I[f,f] \,,
\ee
is called the Enskog-Boltzmann equation \cite{Ernst98}. In the following, however, we stick to the simple Boltzmann equation for simplicity.

In its present form, Eq.~(\ref{eq:Boltz:gran}) does not display a stationary state, because energy is continuously lost.
One may then study scaling regimes like the so-called homogeneous cooling state  \cite{Haff,Zanetti93,Brey96}.
To compensate for the energy loss, and reach a stationary state, it is necessary to include a driving force.
In an experiment, energy injection is implemented by strongly shaking the container in which the grains are enclosed \cite{Gradenigo11b,Puglisi12,Naert12a,Naert12b}.
Theoretically, it is more convenient, however, to assume that a random driving force is continuously acting on the particles, according to
\be
\frac{d{\bf v}_i}{dt} = \frac{{\bf F}_i(t)}{m} + {\boldsymbol \xi}_i(t)
\ee
where ${\bf F}_i(t)$ is the force generated by collisions involving particle $i$, and ${\boldsymbol \xi}_i(t)$ is the random driving force, modeled as
a white noise,
\be \label{eq:noise:granul}
\la \xi_{i,\alpha}(t) \ra = 0, \qquad
\la \xi_{i,\alpha}(t) \xi_{j,\beta}(t') \ra = 2D \, \delta_{ij} \,
\delta_{\alpha\beta} \, \delta(t-t').
\ee
In this case, the Boltzmann equation is modified by adding a diffusion term
to account for the random force:
\be \label{eq:Boltz:gran2}
\frac{\partial f}{\partial t} + \nabla \cdot ({\bf v} f) = I[f,f]
+ D (\nabla_{\bf v})^2 f.
\ee

\subsubsection{Hydrodynamic equations for the slow fields}

From Eq.~(\ref{eq:Boltz:gran2}), it is possible to derive hydrodynamic equations for the relevant slow fields.
Slow fields are usually determined by conservation laws and, when relevant, by order parameters close to a symmetry breaking transition. These fields evolve on time scales that are much larger than other modes (called fast modes), so that only these slow modes have to be retained in a large scale statistical description.
Here, conserved quantities during collisions (and thus during the whole dynamics of the system) are the number of particles and the total momentum.
Slow fields are thus the density and momentum fields.
For elastic particles, the total energy would also be conserved.
In a granular gas, energy is not conserved due to inelastic collisions.
Yet, one can assume that the restitution coefficient $\alpha$ is close to $1$,
$\alpha=1-\epsilon$ with $\epsilon \ll 1$, in such a way that the relaxation of energy occurs on times much longer than the relaxation time of fast modes, so that the time scale separation still holds. In this case, energy is thus still considered as a slow mode, and included in the large scale description.

In the absence of an external potential, energy reduces to kinetic energy, and the local average kinetic energy is simply called granular temperature.
In addition, since all particles have the same mass $m$, one can use the velocity instead of momentum to define a hydrodynamic field.
The density field has been defined in Eq.~(\ref{eq:density:gran}),
and the hydrodynamic velocity and temperature fields are defined as
\bea
{\bf u}({\bf r},t) &=& \frac{1}{\rho({\bf r},t)} \int d{\bf v} \, {\bf v}
f({\bf r},{\bf v},t) \,,\\
{\bf T}({\bf r},t) &=& \frac{1}{\rho({\bf r},t)} \int d{\bf v} \, \frac{1}{2}m{\bf v}^2 f({\bf r},{\bf v},t) \,.
\eea
The continuity equation for the density field is simply obtained by integrating the Boltzmann equation (\ref{eq:Boltz:gran2}) over the velocity, yielding the standard equation
\be \label{eq:gran:rho}
\frac{\partial \rho}{\partial t} + \nabla \cdot (\rho {\bf u}) = 0.
\ee
To obtain equations for the velocity and temperature fields,
one assumes that the distribution $f({\bf r},{\bf v},t)$ depends on space and time only through the hydrodynamic fields 
$\rho({\bf r},t)$, ${\bf u}({\bf r},t)$ and $T({\bf r},t)$, namely \cite{Puglisi-book}
\be
f({\bf r},{\bf v},t) = \tilde{f}\big({\bf v}-{\bf u}|\rho({\bf r},t),{\bf u}({\bf r},t),T({\bf r},t)\big) \,.
\ee
Using this form, one can derive the following hydrodynamic equations for the velocity and temperature fields \cite{Trizac99}
\bea \label{eq:gran:u}
\frac{\partial {\bf u}}{\partial t} + {\bf u} \cdot \nabla {\bf u} = -\frac{1}{m\rho}
\nabla \cdot {\boldsymbol \Pi} \,,\\
\frac{\partial T}{\partial t} + {\bf u} \cdot \nabla T =
-\frac{2}{d\rho} (\nabla \cdot {\bf J} + {\boldsymbol \Pi}:\nabla {\bf u})
-\Gamma + 2mD \,,
\label{eq:gran:T}
\eea
where $d=2$ or $3$ is the space dimension, ${\boldsymbol \Pi}$ is the pressure tensor, ${\bf J}$ is the heat flux, and $\Gamma$ is an energy loss term related to particle inelasticity;
$D$ is the diffusion coefficient of the noise, defined in Eq.~(\ref{eq:noise:granul}). The pressure tensor is expressed as
\be
\Pi_{\alpha \beta} = p \delta_{\alpha \beta} + \Pi_{\alpha \beta}^{\rm diss}
\ee
where $\Pi_{\alpha \beta}^{\rm diss}$ is the dissipative momentum flux.
The latter can in turn be expressed as a linear function of
$\nabla_{\alpha} u_{\beta}$ \cite{Trizac99}, by introducing the kinematic
viscosity $\nu$ and the longitudinal viscosity $\nu_l$ \cite{Trizac99}.
The heat flux satisfies a Fourier law ${\bf J}=-\kappa\nabla T$,
with $\kappa$ the heat conductivity.
If the inelasticity is small enough, the transport coefficients
$\nu$, $\nu_l$ and $\kappa$ can be estimated from the Enskog theory of elastic hard spheres \cite{Trizac99,Enskog}.

Taken together, Eqs.~(\ref{eq:gran:rho}), (\ref{eq:gran:u}) and (\ref{eq:gran:T}) constitute the granular hydrodynamics. They can be used to study the large scale behaviour of a granular gas, like the steady state of a granular gas in a shaken container and the convection instability of this state or, in the absence of driving ($D=0$), the homogeneous cooling state and its shear and heat instabilities \cite{Puglisi-book}.

As we have seen, conservation laws play an important role in the identification of slow fields. For granular gases, only the number of particles and the total momentum are conserved. However, it is possible to consider models in which these quantities are no longer conserved. An example is the case of probabilistic ballistic annihilation, in which particles either collide elastically or annihilate \cite{Coppex04a,Coppex04b,Coppex05}.
This annihilation process makes the number of particles non-conserved, which in turn implies a non-conservation of momentum and energy.
If, however, the annihilation rate is small, most collisions will be elastic, so that the number of particles, momentum and energy still evolve on time scales that are much larger than the relaxation scale of fast modes.
The density, momentum (or velocity) and temperature fields can thus still be considered as the hydrodynamic modes of the system, due to the time scale separation.

\subsection{Self-propelled particles with aligning binary collisions}
\label{sec-SPP-Boltz}

Systems of self-propelled particles can also be treated in the framework of kinetic theory, notably through the Boltzmann equation. This is especially relevant for the case of aligning interactions, which may be treated as binary collision events in the dilute limit.
This approach has been applied in particular to a Vicsek-like model with binary collisions \cite{BDG06,BDG09,EPJST14,Weber13} ---see also \cite{Vicsek95,Chate04} for the definition of the Vicsek model and \cite{Aranson05} for a seminal contribution to a related problem.

As already discussed in Sect.~\ref{sec-kinth-grangas},
the Boltzmann approach uses the simplest closure, which assumes that
$f_2({\bf r}_1,{\bf v}_1,{\bf r}_2,{\bf v}_2) = f_1({\bf r}_1,{\bf v}_1)
f_1({\bf r}_2,{\bf v}_2)$,
where $f_1$ and $f_2$ are respectively the one-point and two-point phase-space distributions.
This closure may be improved by taking into account a correction factor $\chi$ introduced as $f_2({\bf r}_1,{\bf v}_1,{\bf r}_2,{\bf v}_2) 
= \chi f_1({\bf r}_1,{\bf v}_1) f_1({\bf r}_2,{\bf v}_2)$, leading to a Enskog-Boltzmann equation.
For granular gases, the correction factor $\chi$ depends on the density $\rho$ (see Sect.~\ref{sec-kinth-grangas}), and takes into account some steric effects appearing in the intermediate density regime.
In the context of self-propelled particles with velocity-aligning interactions, it has been proposed that the correction factor $\chi$ should be included even in the low density limit; The factor $\chi$ then depends on the angle difference between the pre-collisional velocity vectors \cite{Hanke13}.

A kinetic theory taking into account multi-particle interactions,
as present in the Vicsek model \cite{Vicsek95}, has also been considered
\cite{Ihle10,Ihle16}.
However, this approach still neglects correlations between particles prior to collisions, assuming a factorized form for the pre-collisional $N$-body distribution. Pre-collisional correlations can be taken into account through 
an involved ``ring-kinetic'' theory, which is based on a 
low-density expansion, through diagrammatic techniques, of the collision operator \cite{Ihle15}.

Coming back to more elementary approaches, let us mention that the Boltzmann equation can also be used for self-propelled particles with nematic interactions \cite{Peshkov12a}, or with polar interactions on a topological neighborhood \cite{Peshkov12b}, as well as in the case of active nematic particles \cite{BertinNJP13,Ngo14}.

\begin{figure}[t!]
\centering\includegraphics[height=4cm]{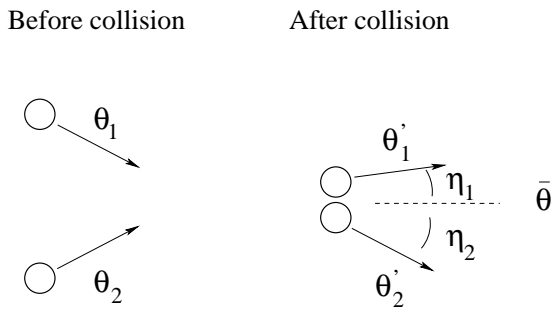}
\caption{Illustration of binary collisions in the model of self-propelled
particles with alignment interactions through binary collisions.
Velocity angles $\theta_i'$ after the collision are equal to the average $\bar\theta$ of the incoming angles $\theta_i$ plus some noise $\eta_i$. Momentum in not conserved in the collision.}
\label{fig-kinth-SPP}
\end{figure}

We consider here a model of particles moving on a two-dimensional plane with a fixed speed $v_0$ in a direction defined by the angle $\theta$.
The angle $\theta$ evolves either through angular diffusion with diffusion coefficient $D_R$, or through a run-and-tumble dynamics, with scattering events $\theta \to \theta + \eta$ (where $\eta$ is a random variable with distribution $p(\eta)$) at a rate $\lambda$ per unit time.
The angle $\theta$ also evolves due to binary collisions which tend to align the velocity angle to the mean direction of motion of the two incoming particles.
Collisions occur when the distance between the two particles is less than the interaction range $d_0$.
After a collision, two particles with incoming angles $\theta_1$ and $\theta_2$ have new angles $\theta_1'$ and $\theta_2'$ given by
\be
\theta_1' = \bar\theta + \eta_1, \qquad \theta_2' = \bar\theta + \eta_2,
\ee
where $\bar\theta$ is the average direction of motion prior to the collision,
\be
\bar\theta = {\rm arg}(e^{i\theta_1} + e^{i\theta_2})
\ee
and where $\eta_1$ and $\eta_2$ are independent random variables drawn from a distribution $p(\eta)$. Note that for simplicity, we have chosen the same noise distribution for the run-and-tumble and collision dynamics, but this needs not be the case.
A sketch of the binary collision is given in Fig.~\ref{fig-kinth-SPP}.

\subsubsection{Boltzmann equation for aligning self-propelled particles}

Similarly to the local mean-field approach described in Sect.~\ref{sec-SPP-FP}, the basic statistical object used in the description is the one-particle phase space distribution $f({\bf r},\theta,t)$.
In the present context, its evolution is governed by
the Boltzmann equation, which reads \cite{BDG06,BDG09}
\be \label{eq-Boltz}
\frac{\partial f}{\partial t}(\mathbf{r},\theta,t) + v_0 \mathbf{e}(\theta)
\cdot \nabla f(\mathbf{r},\theta,t) = I_{\mathrm{dif}}[f]+I_{\mathrm{col}}[f,f].
\ee
The functionals  $I_{\mathrm{dif}}[f]$ and $I_{\mathrm{col}}[f,f]$
respectively account for the angular diffusion (or run-and-tumble) and collision phenomena.
The vector $\mathbf{e}(\theta)$ is the unit vector in the direction $\theta$.
The diffusion functional $I_{\mathrm{dif}}[f]$ is given by
\be
I_{\mathrm{dif}}[f] = D_R \frac{\partial^2 f}{\partial \theta^2}
\ee
in the case of angular diffusion, and by
\be
I_{\mathrm{dif}}[f] = -\lambda f(\mathbf{r},\theta,t)
+ \lambda \int_{-\infty}^{\infty} d\eta\, p(\eta) f(\mathbf{r},\theta-\eta,t)
\ee
in the case of run-and-tumble dynamics.
To evaluate  the collision term $I_{\mathrm{col}}[f,f]$, one needs to consider the details of the collision dynamics.
Two particles collide if their distance becomes less
than the interaction range $d_0$.
In the frame of particle $1$, particle $2$ has a velocity
$\mathbf{v}_2'=v_0 [\mathbf{e}(\theta_2)-\mathbf{e}(\theta_1)]$. Hence,
particles that collide with particle $1$ between $t$ and $t+dt$ are those
that lie, at time $t$, in a rectangle of length $|\mathbf{v}_2'|\, dt$ and
of width $2d_0$, yielding for the collision functional \cite{BDG06,BDG09}
\bea
I_{\mathrm{col}}[f,f] &=&  - 2d_0v_0 f(\mathbf{r},\theta,t) \int_{-\pi}^{\pi} d\theta'\,
|\mathbf{e}(\theta')-\mathbf{e}(\theta)| f(\mathbf{r},\theta',t)\\
\nonumber
&+& 2d_0v_0 \int_{-\pi}^{\pi} d\theta_1 \int_{-\pi}^{\pi} d\theta_2
\int_{-\infty}^{\infty} d\eta\, p(\eta)\,
|\mathbf{e}(\theta_2)-\mathbf{e}(\theta_1)|\\
\nonumber
&& \qquad \qquad \qquad \qquad \times f(\mathbf{r},\theta_1,t) f(\mathbf{r},\theta_2,t)
\delta_{2\pi}(\overline{\theta}+\eta-\theta),
\eea
with $\overline{\theta}=\arg(e^{i\theta_1}+e^{i\theta_2})$,
and $\delta_{2\pi}$ is a generalized Dirac distribution taking into account the periodicity of angles.
Here again, as in the local mean-field approach, the uniform distribution 
$f(\mathbf{r},\theta)=\rho/2\pi$ is a steady-state solution of
Eq.~(\ref{eq-Boltz}) for an arbitrary constant density $\rho$,
and for any noise distributions $p(\eta)$.

In spite of some formal similarities, the Boltzmann equation Eq.~(\ref{eq-Boltz}) and the local mean-field equation Eq.~(\ref{eq:MFFP}) have different interpretations.
In the local mean-field case, particles have a continuous interaction that lasts all the time they are within the interaction range.
The latter is supposed to be sufficiently large to include a significant number of particles to justify the introduction of an average force, but should be small enough to avoid significant variations of $f(\mathbf{r},\theta,t)$ over this range. In the Boltzmann approach, one instead treats interactions as instantaneous binary collisions. Hence the collision term in Eq.~(\ref{eq-Boltz}) depends on the probability of collision, which is a function of the velocity angles of the two incoming particles.
In contrast, in the local mean-field approximation, one takes into account the probability to be within the interaction range and averages the force over this probability, thus discarding information on the velocity of particles.

\subsubsection{Hydrodynamic equations for density and order parameter fields}

As we have seen in Sect.~\ref{sec-SPP-FP}, the relevant slow fields to be kept in the hydrodynamic description are the density field, due to the conservation of the number of particles, and the velocity (or momentum) field, due to the presence of a symmetry breaking. Note that here, momentum appears in the hydrodynamic description as an order parameter, which has a slow dynamics close to the transition, and not as a conserved quantity as in granular gases.
The procedure used to derive hydrodynamic equations for the density and velocity field follows closely the one presented for the local mean-field approximation in Sect.~\ref{sec-SPP-FP}. One starts by expanding the Boltzmann equation into angular Fourier modes \cite{BDG09}:

\be
\frac{\partial{\hat{f}_k}}{\partial t} + \frac{v_0}{2} ( \hat{\triangledown} \hat{f}_{k-1} + \hat{\triangledown}^* \hat{f}_{k+1} ) = - D_k \hat{f}_k + 
2d_0v_0 \sum_{q=-\infty}^{\infty} (\hat{P}_k I_{q-k/2}-I_q) \hat{f}_q \hat{f}_{k-q}
\label{eq-fk}
\ee
with $D_k = D_R k^2$ for angular diffusion and
$D_k = \lambda (1-\hat{P}_k)$ for run-and-tumble dynamics,
where $\hat{P}_k = \int_{-\infty}^{\infty} d\eta\, p(\eta)\, e^{ik\eta}$
(for a Gaussian noise distribution, $\hat{P}_k=e^{-\sigma^2 k^2/2}$).
The complex differential operators $\hat{\triangledown}$ and
$\hat{\triangledown}^*$ have been defined in Eq.~(\ref{eq:def:nabla:complex}).
The coefficient $I_k$ in Eq.~(\ref{eq-fk}) is defined by the integral
\begin{equation}
I_k = \frac{1}{\pi} \int_{-\pi}^{\pi} d\theta \, \left|\sin \frac{\theta}{2} \right| \cos(k\theta).
\end{equation}
The continuity equation, obtained from Eq.~(\ref{eq-fk}) for $k=0$
(recalling that $\hat{f}_0=\rho$),
is identical to Eq.~(\ref{eq:contin:SPP}) since it only characterizes the conservation of the number of particles, and does not depend on the details of the interactions.

The equation for the velocity field ${\bf v}$ (or, in practice, the `momentum' field ${\bf w} = \rho {\bf v}$), is obtained in the same way as in the local mean-field approximation. This comes from the fact that Eqs.~(\ref{eq:MFFP:Fourier}) and (\ref{eq-fk}) share a very similar structure.
One then uses the scaling ansatz given in Eq.~(\ref{eq:scaling:SPP}), and truncates the equation for $\hat{f}_1$ and $\hat{f}_2$ to order $\epsilon^3$.
Then $\hat{f}_2$ is slaved to $\hat{f}_1$, which leads to a closed nonlinear equation involving $\hat{f}_1$ and $\rho$.
Mapping this equation to two-dimensional vectors yields the following evolution
equation for the field 
\bea \label{eq-w:boltz}
\frac{\partial \mathbf{w}}{\partial t} + \gamma (\mathbf{w} \cdot \nabla)
\mathbf{w} &=& -\frac{1}{2}\nabla (\rho - \kappa \mathbf{w}^2) \\
\nonumber
&+& (\mu - \xi \mathbf{w}^2) \mathbf{w}
+ \nu \Delta \mathbf{w} - \kappa (\nabla \cdot \mathbf{w}) \mathbf{w}.
\eea
This equation, derived in the Boltzmann framework, precisely takes the same form as Eq.~(\ref{eq-w}) which was derived from the local mean-field approximation,
but with a different expression of the coefficients as a function of the microscopic parameters and of the local density \cite{BDG06,BDG09}.
Both Eqs.~(\ref{eq-w}) and (\ref{eq-w:boltz})
correspond to the (noiseless) Toner-Tu equation \cite{TT95,TT98,TT12,TTR-rev}, that had been originally obtained in a phenomenological way from symmetry arguments. 
Indeed, Eqs.~(\ref{eq-w}) and (\ref{eq-w:boltz}) include all the terms allowed by the symmetry of the problem, up to order $\epsilon^3$ according to the scaling ansatz Eq.~(\ref{eq:scaling:SPP}).
It is thus not surprising that both the local mean-field and the kinetic theory approach lead to equations of the same form.
The advantage of deriving Eq.~(\ref{eq-w:boltz}) from a
microscopic model is, whatever the method employed, that the coefficients appearing in the equation are known as a function of a reduced number of microscopic parameters. Such an approach thus allows for an easier study of the 'physical' phase diagram of the model (see, e.g., \cite{EPJST14}).

Although the coefficients obtained from the Boltzmann and local mean-field approaches have different expressions, the resulting phase diagram is qualitatively the same \cite{CaussinPRL15,CaussinPRE15}.
A key feature in both cases is that the linear coefficient $\mu$ appearing
in Eq.~(\ref{eq-w:boltz}) changes sign as a function of both density $\rho$ and noise amplitude $D_1$, thus defining a critical line $\rho_c(D_1)$
---we recall that $D_1=D_R$ for angular diffusion, and
$D_1=\lambda(1-\hat{P}_1)$ for run-and-tumble.

The coefficient $\mu$ is negative for $\rho<\rho_c(D_1)$,
and positive for $\rho>\rho_c(D_1)$.
As a first consequence, the state ${\bf w}=0$ is unstable for $\rho>\rho_c(D_1)$, and a bifurcated state characterized by $||{\bf w}||=\sqrt{\mu/\xi}$ emerges, corresponding to the onset of collective motion.
However, it has been shown that due to the density dependence of coefficient $\mu$, this homogeneous state of collective motion is unstable to finite wavelength perturbations along the direction of order (compression waves) \cite{BDG09}
 when the average density $\rho$ is close to $\rho_c$ (with $\rho>\rho_c$).
This leads, within a given range of parameters, to the onset of solitary waves
of high density moving over a disordered, low density background
\cite{BDG09,CaussinPRL15,CaussinPRE15}, in agreement with numerical simulations of the Vicsek model \cite{Chate04,Chate08}.
Further away from the transition, at higher density or lower noise, the stability of the homogeneous state of collective motion is recovered \cite{EPJST14}.
At even lower noise, a second instability occurs, but it is considered as
'spurious' in the sense that it is not observed in numerical simulations of particle-based (Vicsek) model. This instability is likely to be related to the truncation procedure, valid only close to the transition.
Note that adding positional diffusion in the model reduces the range of parameters where this secondary instability is present \cite{EPJST14}.

A further issue concerns the validity of the 'molecular chaos' hypothesis
done in kinetic theory, that is the assumption that particles are decorrelated
before collisions.
Numerical tests of this hypothesis of have been performed in particle-based models, indicating that corrections to the standard Boltzmann equation should be taken into account even in the limit of low density \cite{Hanke13}; The Boltzmann equation can then be replaced by an Enskog-Boltzmann equation as mentioned in the introduction of this subsection.
This comes from the fact that, at low density, the transition to collective motion occurs for a low intensity of the collision noise.
This implies that the velocity of colliding particles are almost identical after a collision, so that the recollision probability may not be negligible.


\section{Determination of the $N$-body distribution}
\label{sec-Nbody}

Up to now, we have discussed different approximation methods (mean-field approximations or kinetic theory) to statistically describe a single particle or elementary unit, interacting with the rest of the system.
One may wonder whether it is possible to find exact solutions for the full joint probability distribution of the $N$ particles or units composing the system.
This is relatively easy in fully-connected models, where the joint probability distribution factorizes (in the large $N$ limit) as a product of $N$ one-body probability distributions
---this is why fully-connected models are considered as mean-field models.
For finite-dimensional geometries, exact solutions of models of interacting dissipative particles (or more general degrees of freedom) are scarce.
Spin-models on a semi-infinite chain with random driving of the first spin \cite{Farago05,Farago07,Farago08}, as well as lattice models with continuous energy transfer and dissipation \cite{Bertin05,Bertin06,DauchotPRL09}, have been proposed.

More generally, models of driven dissipative systems may also be considered from the point of view of their ``natural'' modes: Fourier modes of the velocity field in fluid dynamics \cite{Frisch}, normal modes in jammed particle systems \cite{Henkes12}, etc. From this perspective, the driving (i.e., energy injection) occurs on scales different from that of energy dissipation.
For instance, energy is injected at large scales in three-dimensional turbulence, and dissipated at small scales. In contrast, energy is injected
at intermediate scales in two-dimensional turbulence, and dissipated at large scales. Whatever the direction of energy flow, one may take advantage of this scale separation to consider very simplified models of energy transfer
between an injection source and a dissipation mechanism.
This simple picture thus allows for the use of standard models
like the Zero Range Process (ZRP) or the Simple Exclusion Process (SEP)
to provide exact solutions for this type of driven dissipative systems.
A brief discussion of such one-dimensional exactly solvable models is given in Sect.~\ref{sec-exact-solv}.

Since exactly solvable models are the exception rather than the rule,
it is most often necessary to rely on approximation schemes to evaluate the
$N$-body distribution in driven dissipative systems.
These approximations may be of several types.
We discuss below two different examples:
the Edwards approach for gently driven dense granular matter 
(or more generally, driven athermal systems with dry friction)
in Sect.~\ref{sec:Edwards-Nbody}, and
an approximate treatment of the $N$-body Fokker-Planck equation for systems
of self-propelled particles 
(described by Langevin equations with coloured noise)
in Sect.~\ref{sec:SPP-Nbody}.

\subsection{Exactly solvable models}
\label{sec-exact-solv}

To discuss one-dimensional exactly solvable models, we start in Sect.~\ref{sec-ZRP} with the simple case of the Zero Range Process (ZRP) where a factorized solution can be found. We then further consider more complicated models like
the Asymmetric Simple Exclusion Process (ASEP) in Sect.~\ref{sec:MPA:ASEP}
and a related reaction-diffusion model in Sect.~\ref{sec:MPA:react-diff}.

\subsubsection{Factorizable case: boundary driven Zero Range Process}
\label{sec-ZRP}

In the ZRP, particles residing on the nodes of a lattice are stochastically transferred from one node to a neighboring one, with a transition rate that depends only on the number of particles on the departure site
\cite{Evans-rev00,Evans-rev05,Harris05,Noh05} ---see also \cite{Zia04a,Zia04b,Majumdar05} for similar models defined with continuous masses instead of discrete particles.
Note that for convenience, we stick here to the usual term of ``particles'' to describe the ZRP, but in the spirit of the driven dissipative dynamics described in this review, we may rather interpret these ``particles'' as fixed amounts of energy that are transferred between different degrees of freedom.

\begin{figure}[t!]
\centering\includegraphics[width=8cm]{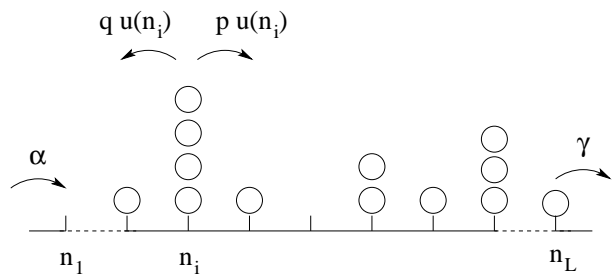}
\caption{Illustration of the ZRP. Particles on site $i$ randomly move to site $i+1$ with probability $pu(n_i)$ and to site $i-1$ with probability $qu(n_i)$.
Particles are also injected on site $i=1$ with rate $\alpha$, and withdrawn from site $L$ with rate $\gamma$.}
\label{fig-ZRP}
\end{figure}

For concreteness, we focus here on a specific one-dimensional version of the
ZRP, with open boundaries (see Fig.~\ref{fig-ZRP}). However, note that exact solutions exist in higher dimensions for ZRP which are not connected to reservoirs \cite{Evans-rev05}.
A boundary-driven ZRP on a tree geometry has also been solved \cite{DauchotPRL09}.

The one-dimensional lattice is thought of as a schematic representation of scale space (for instance of Fourier modes of some relevant physical field); the left boundary represents the largest scale in the system, while the right boundary represents the smallest scale.
The stochastic rules at the boundaries aim at modeling the injection of energy at large scale and the dissipation at small scale. One could of course also exchange the role of 'large' and 'small' scales in the system, to model an inverse cascade.

The transfer dynamics within the system is defined as follows.
The $L$ sites of the lattice are labelled by an index $i=1,\dots,L$,
and the number of particles on site $i$ is denoted as $n_i$.
A particle on site $i$ is transferred to site $i+1$ (if $i+1 \le L$) with a probability $p u(n_i)$ per unit time, and to site $i-1$ (if $i-1 \ge 1$) with a probability $q u(n_i)$ per unit time. Here, $p$ and $q$ are two positive numbers such that $p\ge q$, and $u(n_i)$ is an arbitrary positive function of the local number $n_i$ of particles on the departure site $i$.
In addition, particles are injected on the left boundary with a probability rate $\alpha$, and are withdrawn at the right boundary with a probability rate $\gamma$. 

A microscopic configuration of the ZRP is given by the set
$\mathcal{C}=(n_1,\dots,n_L)$ of the occupation numbers of all sites.
The transition rate $W(\mathcal{C}'|\mathcal{C})$ can be written formally as
\bea \nonumber
W(\{n_i'\}|\{n_i\}) &=& \sum_{i=1}^{L-1} p u(n_i)\,
\delta_{n_i',n_i-1}\, \delta_{n_{i+1}',n_{i+1}+1}
\prod_{j \ne i,i+1} \delta_{n_j',n_j}\\ \nonumber
&+& \sum_{i=2}^L q u(n_i)\,
\delta_{n_i',n_i-1}\, \delta_{n_{i-1}',n_{i-1}+1}
\prod_{j \ne i,i-1} \delta_{n_j',n_j}\\
&+& \alpha \delta_{n_1',n_1+1} \prod_{j=2}^L \delta_{n_j',n_j}
+ \gamma u(n_L) \delta_{n_L',n_L-1} \prod_{j=1}^{L-1} \delta_{n_j',n_j}
\label{eq:transrate:ZRP}
\eea
where $\delta_{n',n}$ is the Kronecker symbol, equal to $1$ if $n'=n$,
and to $0$ otherwise.
Using this form of the transition rate, one can write the corresponding
master equation for the joint probability distribution $P(n_1,\dots,n_L)$, that formally reads
\be
\frac{\partial P}{\partial t}(\{n_i\},t)
= \sum_{\{n_i'\}} [W(\{n_i\}|\{n_i'\}) P(\{n_i'\},t)
- W(\{n_i'\}|\{n_i\}) P(\{n_i\},t)] \,.
\ee
It can be shown \cite{Schutz05} that the steady-state distribution
takes a factorized form
\be \label{eq:dist:ZRP}
P(\{n_i\}) = \prod_{i=1}^L p_i(n_i)
\ee
where the marginal distribution $p_i(n_i)$ reads
\be
p_i(n_i) = \frac{z_i^{n_i}}{Z_i} \, \prod_{k=1}^{n_i} \frac{1}{u(k)}
\ee
with $Z_i$ a normalization constant.
The fugacity $z_i$ is given in terms of the model parameters as \cite{Schutz05}
\be
z_i = \frac{\alpha}{p-q} + \frac{\alpha(p-q-\gamma)}{\gamma(p-q)} \left( \frac{p}{q} \right)^{L-i}
\ee
for $p>q$ and as
\be \label{def-zi-ZRP}
z_i = \frac{\alpha}{\gamma}+\alpha(L-k)
\ee
for $p=q$ (we assume in Eq.~(\ref{def-zi-ZRP}) that $p=q=1$ without loss of generality).
The current $J$ of particles can be easily computed from these expressions,
and reads for all $p \ge q$ simply as $J=\alpha$.
Hence the current is completely controlled by the injection mechanism.

The factorized solution Eq.~(\ref{eq:dist:ZRP}) makes the study of the properties of the model relatively straightforward. Among the standard properties of ZRP is the condensation transition, whereby a finite fraction of the mass present in the system condenses on a single site \cite{Evans-rev05}.
However, this phenomenon requires to consider an isolated ZRP,
where the total number of particles is fixed, at variance with the boundary driven case considered here.
Closer to the model presented here, a boundary driven ZRP
on a tree geometry has been shown to exhibit a transition, as a function of a parameter controling the internal energy transfer, between a state of vanishing flux of dissipated energy in the infinite size limit and a state where this flux remains finite \cite{DauchotPRL09}.

\subsubsection{Matrix Product Ansatz for the open ASEP model}
\label{sec:MPA:ASEP}

In the above example, the joint distribution $P(\{n_i\})$ was taking a very simple factorized form as given in Eq.~(\ref{eq:dist:ZRP}), even though the state of the system is heterogeneous, with a one-site probability distribution $p_i(n_i)$ that explicitly depends on site $i$.
In most cases, however, even simple one-dimensional models do not have an exact solution in terms of a factorized joint probability distribution.
This is the case for instance when constraints on the maximal number of particles per site are taken into account, as in the ASEP, where at most one particle can occupy a given site.
A similar situation occurs when reactions (in the sense of reaction-diffusion processes) are included.
In both cases, exact solutions have been found in terms of the so-called Matrix Product Ansatz (an ansatz inspired by the formalism of quantum mechanics), as briefly described below.
Note that the Matrix Product Ansatz can also be used to solve more general versions of the ASEP model, for instance by including two species of particles \cite{Prolhac09,Ragoucy15,Ragoucy16}.

\begin{figure}[t!]
\centering\includegraphics[width=8cm]{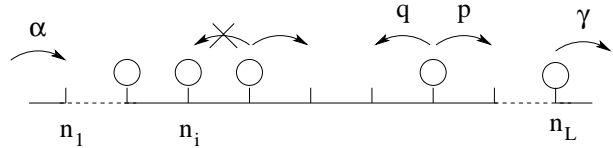}
\caption{Illustration of the ASEP. Particles on site $i$ randomly jump to site $i+1$ with probability $p$ and to site $i-1$ with probability $q$, provided that the target site is empty (otherwise the transition is forbidden).
Particles are also injected on site $i=1$ with rate $\alpha$, and withdrawn from site $L$ with rate $\gamma$.}
\label{fig-ASEP}
\end{figure}

In the ASEP, at most one particle can sit on each site of a one-dimensional lattice of $L$ sites.
Hence the number of particles $n_i$ on site $i$ equals $0$ or $1$.
A particle $i$ can move to site $i+1$ (resp.~$i-1$) with probability rate $p$
(resp.~$q$), on condition that the target site is empty
(see Fig.~\ref{fig-ASEP}).
As in the case of the ZRP, we focus on the specific situation where particles are injected at rate $\alpha$ on the site $i=1$, and withdrawn at rate $\gamma$ from site $i=L$, thus again modeling injection and dissipation of energy at different scales, but here with a limitation on the transfer capacity.
In most papers on the ASEP model, $p$ and $q$ are not kept as independent parameters, but one 'normalizes' the time scale by setting $p=1$, $pq=1$ or $p+q=1$ for instance. However, note that the way $p$ and $q$ are normalized
has an influence on the definition of the injection and dissipation rates $\alpha$ and $\gamma$.

The transition rate $W(\mathcal{C}'|\mathcal{C})$ now reads
\bea \nonumber
W(\{n_i'\}|\{n_i\}) &=& \sum_{i=1}^{L-1} p\,
\delta_{n_i',n_i-1}\, \delta_{n_{i+1}',n_{i+1}+1} \delta_{n_{i+1},0}
\prod_{j \ne i,i+1} \delta_{n_j',n_j}\\ \nonumber
&+& \sum_{i=2}^L q \,
\delta_{n_i',n_i-1}\, \delta_{n_{i-1}',n_{i-1}+1} \delta_{n_{i-1},0}
\prod_{j \ne i,i-1} \delta_{n_j',n_j}\\
&+& \alpha \delta_{n_1',n_1+1} \delta_{n_1,0} \prod_{j=2}^L \delta_{n_j',n_j}
+ \gamma \delta_{n_L',n_L-1} \prod_{j=1}^{L-1} \delta_{n_j',n_j}
\label{eq:transrate:SEP}
\eea
These transition rates share some similarities with the transition rates of the Zero Range Process given in Eq.~(\ref{eq:transrate:ZRP}), but the main difference is the presence in Eq.~(\ref{eq:transrate:SEP})
of additional Kronecker delta symbols that enforce that the target site has to be empty when moving or adding a particle.
The rate $u(n_i)$ that appears in Eq.~(\ref{eq:transrate:ZRP}) would reduce in the ASEP to $u(1)$, which can be reabsorbed into the definition of the rate $p$ and $q$, and thus does not appear explicitly in Eq.~(\ref{eq:transrate:SEP}).

The master equation associated with the transition rates (\ref{eq:transrate:SEP}) can be solved in steady-state using a Matrix Product Ansatz.
The basic idea is to generalize the simple product structure of the Zero Range Process, as given in Eq.~(\ref{eq:dist:ZRP}) into a product of matrices
\be \label{eq:MPA}
P(\{n_i\}) = \frac{1}{Z} \mathcal{L}\big( M(n_1) M(n_2)\dots M(n_L) \big)
\ee
where $M(n)$ is a matrix-valued function, and $\mathcal{L}$ is a linear operator that projects matrices onto real values.
In practice, $\mathcal{L}$ usually takes the form
\be
\mathcal{L}(M) = \la W|M|V \ra
\ee
for an open system, with $\la W|$ and $|V \ra$ two vectors
(notations are borrowed from quantum mechanics), and
$\mathcal{L}(M) = {\rm Tr}(M)$ for systems with periodic boundaries (this case will not be described further here).
The normalization constant $Z$ takes a simple formal expression,
\be
Z = \la W| C^L |V \ra, \qquad C \equiv \sum_{n} M(n).
\ee
Interestingly, a proof that the $N$-body stationary distribution takes a Matrix Product form like Eq.~(\ref{eq:MPA}) has been given 
for one-dimensional lattice models with local stochastic rules (e.g., short range jumps)
\cite{Sandow97}.
However, this existence proof does not provide in itself the way to determine explicitly the matrix $M(n)$, and thus the distribution.

In the case of the ASEP, $n_i$ takes only two values $0$ and $1$, so that the determination of the function $M(n)$ boils down to the determination of two matrices $D\equiv M(1)$ and $E \equiv M(0)$.
By using the ansatz (\ref{eq:MPA}) in the master equation, one finds that the matrices $D$ and $E$ need to satisfy the following algebraic relation
(assuming $p=1$ without loss of generality) \cite{Mallick97}
\be \label{eq:commut:SEP}
DE-qED=(1-q)(D+E).
\ee
The vectors $\la W|$ and $|V \ra$ should in addition satisfy the following relations, deduced from the injection and dissipation rates \cite{Mallick97},
\be \label{eq:boundary:vectors:SEP}
\la W|E = \frac{1-q}{\alpha} \la W|, \qquad 
D |V\ra = \frac{1-q}{\gamma} |V\ra \,.
\ee
Explicit matrices $D$, $E$ and vectors $\la W|$ and $|V\ra$ satisfying
Eqs.~(\ref{eq:commut:SEP}) and (\ref{eq:boundary:vectors:SEP}) can be found
\cite{Derrida93,Sandow94,Essler96,Mallick97,Essler00,Evans-MPArev}.
In most cases, it is necessary to use infinite dimensional matrices and vectors to fulfill 
Eqs.~(\ref{eq:commut:SEP}) and (\ref{eq:boundary:vectors:SEP}).
However, finite dimensional representations exist when a specific relation is imposed between the parameters $p$ and $q$ and the boundary exchange rates with the reservoirs \cite{Mallick97}.
In the case $p=1$ and $q=0$, in which case only infinite representations exist, a possible choice of the matrices $D$ and $E$ is given by \cite{Derrida93}
\be
D =
\begin{pmatrix}
 1 & 1 & 0 & 0 & \cdots \\
 0 & 1 & 1 & 0 & \\
 0 & 0 & 1 & 1 & \\
 0 & 0 & 0 & 1 & \\
\vdots & & & & \ddots \\
\end{pmatrix}
, \qquad
E =
\begin{pmatrix}
 1 & 0 & 0 & 0 & \cdots \\
 1 & 1 & 0 & 0 & \\
 0 & 1 & 1 & 0 & \\
 0 & 0 & 1 & 1 & \\
\vdots & & & & \ddots \\
\end{pmatrix}
\ee
with boundary vectors $\la W|$ and $|V \ra$
\be
\la W| = \kappa
(1, a, a^2, a^3, \dots), \qquad
|V \ra = \kappa (1, b, b^2, b^3, \dots)^T
\ee
where the parameters $a$, $b$ and $\kappa$ are defined as
\be
a = \frac{1-\alpha}{\alpha}, \quad
b = \frac{1-\gamma}{\gamma}, \quad
\kappa = \sqrt{\frac{\alpha+\gamma+1}{\alpha \gamma}} \,.
\ee
This choice of $D$, $E$, $\la W|$ and $|V \ra$ has the advantage of being mathematically elegant, in the sense that boundary rates $\alpha$ and $\gamma$ appear only in the boundary vectors $\la W|$ and $|V \ra$ (this is not the case for all representations of $D$, $E$, $\la W|$ and $|V \ra$), and that $D$ and $E$ have a similar form which emphasizes the particle-hole symmetry present in the model. In practice, however, this choice leads to divergences in some parameter range when computing correlation functions, so that other representations may be preferred \cite{Derrida93}.

The matrix product form of the stationary distribution is convenient to compute the average density profile and two-point (or multi-point) correlation functions. 
For instance, the average density $\la n_i \ra$ on site $i$ is obtained by summing $P(n_1,\dots,n_L)$ over all $n_j$ with $j \ne i$, yielding
\be
\la n_i \ra = \frac{\la W| C^{i-1} D C^{L-i} |V \ra}{\la W| C^L |V \ra} \,,
\ee
with $C=D+E$.
Similarly, the two-point correlation function $\la n_i n_j \ra$
is obtained for $j>i$ as
\be
\la n_i n_j \ra = \frac{\la W| C^{i-1} D C^{j-i-1} D C^{L-j} |V \ra}{\la W| C^L |V \ra}.
\ee
The average current $J$ accross a given link $(i,i+1)$ also takes a simple formal expression (note that $J$ is uniform throughout the system in steady state, and thus independent of $i$). The current $J$ is defined as
\be
J = \la n_i(1-n_{i+1}) - q(1-n_i)n_{i+1} \ra
\ee
(we recall that $p$ has been set to $1$), leading, using the matrix product form, to
\be
J = \frac{\la W| C^{i-1} (DE-qED) C^{L-j-1} |V \ra}{\la W| C^L |V \ra} \,.
\ee
From the algebraic relation (\ref{eq:commut:SEP}), the expression of the current simplifies to
\be \label{eq:J:MPA}
J = (1-q) \frac{\la W| C^{L-1} |V \ra}{\la W| C^L |V \ra}.
\ee
In practice, the quantity $\la W| C^L |V \ra$ often takes at large $L$ the generic form \cite{Derrida93}
\be
\la W| C^L |V \ra \sim A L^z \lambda^L
\ee
where $A$, $z$ and $\lambda$ do not depend on the system size $L$.
One then finds by taking the limit $L \to \infty$ in Eq.~(\ref{eq:J:MPA}) that
$J=(1-q)/\lambda$.

The phase diagram of the ASEP model can be deduced from the evaluation 
of the current $J$. For simplicity, we discuss here only the case of a totally asymmetric dynamics ($q=0$).
One finds three phases \cite{Derrida93}: (i) a maximal-current phase
$(\alpha > \frac{1}{2},
\gamma > \frac{1}{2})$ for which $J=\frac{1}{4}$,
(ii) a low-density phase $(\alpha < \frac{1}{2}, \gamma > \alpha)$
where the current $J=\alpha(1-\alpha)$
is controlled by the injection reservoir, and (iii)
a high density phase $(\gamma < \frac{1}{2}, \gamma < \alpha)$
in which the current $J=\gamma(1-\gamma)$
is controlled by the dissipation mechanism.
The line $\alpha=\gamma<\frac{1}{2}$ is a first order phase transition
\cite{Derrida93}.
The phase diagram remains qualitatively the same for $q>0$ \cite{Sandow94}.
One can also show from the Matrix Ansatz formulation that
long-range correlations, of weak amplitude proportional $L^{-1}$,
but spanning the whole system size, are present in the model
\cite{Derrida07}.

\subsubsection{Matrix Product solution for reaction-diffusion models}
\label{sec:MPA:react-diff}

We have discussed above the ZRP and ASEP models, having in mind an interpretation in terms of transfer of energy between scales.
Coming back to a particle picture, the irreversibility of the dynamics may
also come from the presence of reactions, in the sense of reaction-diffusion processes. Some one-dimensional reaction-diffusion processes can also be solved using the Matrix Product Ansatz method \cite{Jafarpour03,Jafarpour04,Sasamoto04,Mohanty09,Jafarpour13}.
This is the case in particular for the so-called ``anisotropic decoagulation model'' \cite{Hinrichsen00}, defined by the following rates:
\bea
\label{eq:RD:diff}
&& \emptyset A \underset{q}{\rightarrow} A \emptyset, \qquad \;
A \emptyset \underset{p}{\rightarrow} \emptyset A \qquad \,
(\rm diffusion)\\
\label{eq:RD:coag}
&& AA \underset{q}{\rightarrow} A \emptyset, \qquad
AA \underset{p}{\rightarrow} \emptyset A \qquad
(\rm coagulation)\\
\label{eq:RD:decoag}
&& \emptyset A \underset{\kappa q}{\rightarrow} A A, \qquad
A \emptyset \underset{\kappa p}{\rightarrow} A A \qquad
(\rm decoagulation)
\eea
with the constraint $pq=1$.
Note that boundaries are closed, in the sense that there is no exchange of particles with external reservoirs at sites $i=1$ and $i=L$
($\alpha=\gamma=0$ in the notations of Sect.~\ref{sec:MPA:ASEP}).
The stationary distribution of the model can be evaluated through the Matrix Product Ansatz (\ref{eq:MPA}), with again $M(1)\equiv D$ and $M(0)\equiv E$.

The generalization of the algebraic relation (\ref{eq:commut:SEP})
to the anisotropic decoagulation model is more complicated, and involves two auxiliary matrices ${\bar D}$ and ${\bar E}$ that do not appear in the expression of the probability distribution, but are required to obtain the necessary cancellations of terms in the master equation. One thus ends up with a set of four algebraic relations involving the matrices $D$, $E$, ${\bar D}$ and ${\bar E}$
\cite{Hinrichsen96,Hinrichsen00}.
A four-dimensional representation of the matrices $D$ and $E$ has been found
\cite{Hinrichsen96}, and is given by
\be
D =
\begin{pmatrix}
 0 & 0 & 0 & 0 \\
 0 & 1-\gamma^{-2} & 1-\gamma^{-2} & 0 \\
 0 & 0 & \gamma^2-1 & 0 \\
 0 & 0 & 0 & 0 \\
\end{pmatrix}
, \qquad
E =
\begin{pmatrix}
 p^2 & p^2 & 0 & 0 \\
 0 & \gamma^{-2} & \gamma^{-2} & 0 \\
 0 & 0 & 1 & q^2 \\
 0 & 0 & 0 & q^2 \\
\end{pmatrix}
\ee
where $\gamma^2 \equiv 1+\kappa$.
The boundary vectors $\la W|$ and $|V \ra$ are given by
\be
\la W| = (1-q^2, \, 1, \, 0, \, a_L), \qquad
|V \ra = (b_L, \, 0, \, q^2, \, q^2-1),
\ee
where the parameters $a_L$ and $b_L$ a priori depend on the system size $L$,
and need to satisfy a given constraint \cite{Hinrichsen96}.
From the matrix product form of the probability distribution, it is possible to evaluate the average density profile, as well as correlation functions.
Two different phases are obtained depending on the value of $\kappa$.
When $\kappa < q^2-1$, a low-density phase is observed, while for
$\kappa > q^2-1$, the stationary state corresponds to a high-density phase
\cite{Hinrichsen00}.
In addition, algebraic long-range correlations are present at the critical
point $\kappa_c = q^2-1$ \cite{Hinrichsen00}.

It may seem surprising that a solution with finite matrices could be found for this reaction-diffusion model, while the solution of the ASEP involves infinite matrices.
As mentioned in Sect.~\ref{sec:MPA:ASEP}, solutions with finite matrices exist for the ASEP provided that specific conditions are imposed on the parameters of the model \cite{Mallick97}.
The situation is to some extent similar in the anisotropic decoagulation model.
Specific relations are imposed between the parameters, since the left and right diffusion rates (\ref{eq:RD:diff}) are equal to the corresponding coagulation rates (\ref{eq:RD:coag}), and the decoagulation rates (\ref{eq:RD:decoag}) are proportional to the latter.
Although we are not aware of explicit results on a more general version of the decoagulation model, it is likely that the matrix product solution (which should exist since dynamical rules are local \cite{Sandow97}) would generically involve infinite matrices.

More generally, the applicability of the Matrix Product Ansatz is limited in most cases by the ability to find a tractable representation of the algebraic relations onto which the master equation is mapped.
However, it is possible in some cases to derive the statistical properties of the model from the sole knowledge of the algebraic relations like 
Eqs.~(\ref{eq:commut:SEP}) \cite{Derrida07}.
In addition, connections to techniques used in the context of integrable quantum systems have recently been emphasized for systems with open boundaries \cite{Ragoucy14,Ragoucy15,Ragoucy16}, even including the possibility of bulk annihilation and creation of pairs \cite{Ragoucy16diss}.
Finally, note that other interesting applications of the Matrix Product Ansatz have been found, for instance in the derivation of the statistics of the current of particles \cite{Mallick11,Lazarescu13}, or in the problem of two coupled Kardar-Parisi-Zhang (KPZ) equations \cite{Ferrari13}.


\subsection{Edwards approach for driven athermal systems with dry friction}
\label{sec:Edwards-Nbody}

A class of driven-dissipative systems for which a more systematic approximation method to determine the $N$-body distribution has been devised is the case of driven, dense granular matter that we have already discussed in Sect.~\ref{sec-levine-grains}.
Such systems have the specificity of being athermal and to involve dry (or solid) friction, which is able to exert static tangential forces, at variance with viscous friction.

\subsubsection{Sampling of configurations: the Edwards hypothesis}

The Edwards approach \cite{EO89,ME89,EM94,EG98,BKVS01}
postulates that in a granular system driven in such a way that it periodically relaxes to a mechanically stable configuration, the probability of a configuration $\mC$ (typically the list of the positions of all grains)
with volume $V(\mC)$ and energy $E(\mC)$ is given by
\be
P(\mC) = \frac{1}{Z}\, \exp\left(-\frac{E(\mC)}{T_{\rm eff}}-\frac{V(\mC)}{X}\right) \, \mathcal{F}(\mC),
\label{eq:edmes}
\ee
with $T_{\rm eff}$ the effective temperature, and $X$ the compactivity.
The function $\mathcal{F}(\mC)$ characterizes the mechanical stability of configuration $\mC$: $\mathcal{F}(\mC)=1$ if $\mC$ is mechanically stable, and $\mathcal{F}(\mC)=0$ otherwise.
Note that other forms of the distribution $P(\mC)$, involving the stress in addition to the energy and volume, have also been proposed
\cite{HHC07,HC09,BE09,BJE12,BZBC13,Daniels}.

Alternative formulations of the Edwards hypothesis may not use
Boltzmann factors, but all of them assume that microscopic configurations having the same values of the global constraints (typically volume and/or energy, but also possibly the stress tensor) are equiprobable.
The main difference with respect to usual statistical mechanics
is that the set of accessible configurations is restricted, since only mechanically stable configurations have a nonzero probability, as witnessed by the function $\mathcal{F}(\mC)$ in Eq.~(\ref{eq:edmes}).
The definition of mechanically stable configurations often involves frictional properties, which have no counterpart at equilibrium.

Although the general form of the probability distribution (\ref{eq:edmes}) looks relatively simple, its implementation in situations of interest is not easy, since the practical evaluation of the function $\mathcal{F}(\mC)$ is difficult in most cases, involving complicated correlations between particles.
We discuss below an example in which calculations can be done explicitly, without resorting to mean-field-type approximations as in
Sect.~\ref{sec-levine-grains}.
The calculation illustrates in particular how the transfer operator technique
(a well-known method at equilibrium, at least in its transfer matrix form)
can be used for athermal nonequilibrium systems. We expect this method to be generically applicable to one-dimensional models which can be described by the Edwards probability distribution (\ref{eq:edmes}).

\subsubsection{Statistics of blocked states in a spring-block model}

\begin{figure}[t!]
\centering\includegraphics[width=9cm]{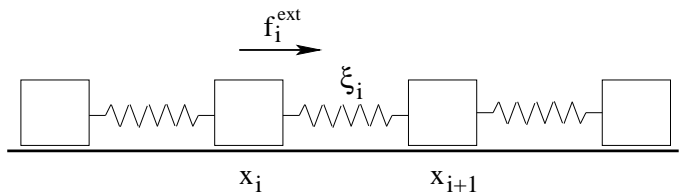}
\caption{Illustration of the spring-block model in the case $N=3$,
showing the position $x_i$ of masses and the elongation
$\xi_i=x_{i+1}-x_i-\ell_0$ of the spring relating masses $i$ and $i+1$. During the driving phase, a force $f_i^{\rm ext}$ is applied on some of the masses.}
\label{fig-spring}
\end{figure}

Let us consider a one-dimensional spring-block model, in which $N+1$ masses are linked by $N$ linear springs of stiffness $k$ 
(see Fig.~\ref{fig-spring}) \cite{Gradenigo15}.
A driving protocol consists of periodic driving cycles, during which
a pulling force is applied to some of the masses for a duration $\tau$, before letting the system relax to a mechanically stable configuration.
Strictly speaking, this model does not describe a granular packing, but it falls within the class of driven athermal systems with dry friction that we are considering in this subsection.

The equation of motion for the position $x_i$ of the $i^{\rm th}$ mass reads
\be\label{eqofmotion}
m \ddot{x}_i = -mg \mu_{\rm d} {\rm sign}(\dot{x}_i) + k(x_{i+1}+x_{i-1}-2x_i) + f_i^{\rm ext},
\ee
where $\mu_{\rm d}$ is the dynamic dry friction coefficient, and $f_i^{\rm ext}$ is the externally applied force (equal to zero in the relaxation phase).
It is convenient to introduce the spring elongation $\xi_i \equiv x_{i+1}-x_i- \ell_0$, where $\ell_0$ is the rest length of the springs.
To avoid crossings of masses, the spring elongations have to satisfy
$\xi_i>-\ell_0$ for all $i$.
Due to the global translational invariance, the configuration of the system can be characterized by the list of spring elongations $(\xi_1,\dots,\xi_N)$, instead of the list of all mass positions $(x_1,\dots,x_{N+1})$; All configurations that differ by a translation are considered as equivalent.

In the absence of driving force, $f_i^{\rm ext}=0$, a configuration $\boldsymbol \xi=(\xi_1,\dots,\xi_N)$ is mechanically stable if
$k|\xi_{i+1}-\xi_i|<\mu_s mg$ for all $i=1,\dots,N-1$.
In the following, we set $k=mg=1$ to lighten notations.
Hence the function $\mathcal{F}(\mC)$ formally defined in Eq.~(\ref{eq:edmes})
can be explicitly written in the present model as
\be
\mathcal{F}(\boldsymbol \xi) = \prod_{i=1}^N \Theta(\mu_{\rm s}-|\xi_{i+1}-\xi_i|),
\ee
where $\Theta$ is the Heaviside function.
Before writing the probability distribution $P(\boldsymbol \xi)$, we first note that the system is not confined (there is no box or confining potential), so that
the ``volume'' of the system (here, its total length) is not directly subjected to a constraint. 
The volume is thus expected not to explicitly appear in Eq.~(\ref{eq:edmes}),
which amounts to having $X=\infty$.
Altogether, the probability of a configuration $(\xi_1,\ldots,\xi_N)$ can be written as
\be
P(\boldsymbol \xi) = \frac{1}{Z}\, e^{-\beta_{\rm eff} \sum_{i=1}^N\xi_i^2/2}\prod_{i=1}^N \Theta(\mu_{\rm s}-|\xi_{i+1}-\xi_i|) \,,
\label{eq:probability_theta}
\ee
where $E=\frac{1}{2}\sum_{i=1}^N\xi_i^2$ is the elastic energy stored in the springs, and $\beta_{\rm eff}=T_{\rm eff}^{-1}$ is the inverse effective temperature.
We recall that $\xi_i>-\ell_0$ for all $i$.

The partition function $Z$, defined by normalization in Eq.~(\ref{eq:probability_theta}), can be evaluated using transfer operator methods.
From Eq.~(\ref{eq:probability_theta}), one has $Z=\Tr(\mT^N)$
where the operator $\mT$ is defined as
$\mT[f](x)=\int_{-\ell_0}^\infty dy\, T(x,y) f(y)$, with $T(x,y)$ the
symmetric kernel:
\be
T(x,y)=e^{-\beta_{\rm eff} x^2/4} \, \Theta(\mu_{\rm s}-|x-y|)\, e^{-\beta_{\rm eff} y^2/4}
\label{eq:transf-op}
\ee
(note that it may be convenient at this stage to take
the limit $\ell_0 \to \infty$ \cite{Gradenigo15}).

Thermodynamic properties of the system can then be determined from the formal expression $Z=\Tr(\mT^N)$ of the partition function $Z$.
In particular, the (effective) free energy $F=-(N\beta_{\rm eff})^{-1} \ln Z$ reduces in the limit $N \to \infty$ to
\be
F = -\frac{1}{\beta_{\rm eff}} \ln \lambda_{\rm max}(\beta_{\rm eff}) \,,
\ee
where $\lambda_{\rm max}(\beta_{\rm eff})$ is the largest eigenvalue  of the operator $\mT$. The average energy per spring $\ve = \langle \frac{1}{2} \xi^2 \rangle$
can be obtained by derivation
\be
\ve = - \frac{\partial \ln \lambda_{\rm max}}{\partial \beta_{\rm eff}}
\ee
(alternatively, one may also compute the energy from the maximal eigenvalue of the operator $\partial \mT/\partial \beta_{\rm eff}$ and the corresponding eigenvector \cite{Gradenigo15}).
The largest eigenvalue $\lambda_{\rm max}$ can be evaluated numerically, by discretizing the operator and using numerical matrix diagonalization procedures \cite{Gradenigo15}.

\subsubsection{Correlation function and infinite temperature critical point}

The transfer operator method can also be used to determine the spatial correlation function of spring elongations, $C_{ij}=\langle \xi_i \xi_j \rangle$.
Numerical simulations of the dynamics \cite{Gradenigo15} show that $C_{ij}$ takes a scaling form
\be
C_{ij} = \tilde{C}\left(\frac{|i-j|}{\ell(\ve)}\right)
\ee
where the correlation length $\ell(\ve)$ turns out to be proportional to the energy density
\be
\ve = \frac{1}{2N} \sum_{i=1}^N \xi_i^2
\ee
of the mechanically stable configuration.
This scaling behaviour of the correlation function $C_{ij}$ and of the correlation length $\ell(\ve)$ is recovered from the transfer operator evaluation
\cite{Gradenigo15}.
It can also be obtained using a simple Gaussian field theory approximation, as follows \cite{Gradenigo15}.
The Heaviside functions appearing in Eq.~(\ref{eq:probability_theta})
can be approximated by Gaussian functions,
\be
\Theta(\mu_{\rm s}-|\xi_{i+1}-\xi_i |) \; \rightarrow \;
\frac{1}{\sqrt{\pi}} \, \exp\left(-\frac{|\xi_{i+1}-\xi_i |^2}{4 \mu_{\rm s}^2}\right),
\ee
leading to
\be
Z \propto \int \mathcal{D}{\boldsymbol\xi} \, e^{-\mathcal{H}(\boldsymbol \xi)}
\ee 
with an effective Hamiltonian given by
\be
\mathcal{H}(\boldsymbol \xi)
= \frac{1}{2}\left[\beta_{\rm eff} \sum_{i=1}^{N} \xi_i^2 + \frac{1}{2\mu_{\rm s}^2}
\sum_{i=1}^N (\xi_{i+1}-\xi_i)^2\right].
\label{eq:eff-act}
\ee
For this (discrete) one-dimensional Gaussian field theory, 
the correlation function is exponential 
\be
\langle \xi_i\xi_j \rangle \sim e^{-|i-j|/\ell}
\ee
with $\ell \sim \sqrt{T_{\rm eff}}$ when $\ell \gg 1$.
As the energy $\ve$ satisfies $\ve \sim \sqrt{T_{\rm eff}}$,
one recovers $\ell \sim \ve$.
This result can also be recovered by taking a continuous limit
in Eq.~(\ref{eq:eff-act}).
From this approach, one sees that the model exhibits an infinite temperature critical point, at odds with standard critical phenomena which in one dimension are found at zero temperature. The reason for this non-standard behaviour can be read off from Eq.~(\ref{eq:eff-act}): the inverse temperature $\beta_{\rm eff}$ appears in front of the 'local' term $\xi_i^2$, while in equilibrium systems it usually appears in front of the 'gradient' term $(\xi_{i+1}-\xi_i)^2$.
This comes from the fact that this gradient term is of entropic nature, resulting from the constraint of mechanical stability, while in equilibrium system, the entropic term is in general local.
In contrast, the energy term which usually takes a gradient form
$(x_{i+1}-x_i)^2$ has been turned to a local term by using the spring extension $\xi_i$ as the relevant variable.

\subsection{Approximate $N$-body distributions for self-propelled particles}
\label{sec:SPP-Nbody}

We have discussed in Sect.~\ref{sec-SPP-FP} and \ref{sec-SPP-Boltz} the case of assemblies of self-propelled particles with dissipative interactions that tend to align the velocities of interacting particles.
Other interesting phenomena like the so-called Motility Induced Phase Separation \cite{Tailleur-rev}, however, occur in systems of self-propelled particles having only conservative repulsive interactions.
The coupling of self-propulsion and repulsion yields in this case an effective attraction, resulting in a phase separation.
The case of self-propelled particles with interaction forces deriving from a potential is thus of interest, and one may wonder if the full $N$-body probability distribution of configurations could be determined, thus generalizing the Boltzmann-Gibbs weight.
Although the full answer to this question is not known, approximate expressions of the $N$-body distribution have been proposed using a Gaussian coloured noise to model the self-propulsion force, and some approximation schemes to derive the distribution. 
These include the Unified Coloured Noise Approximation (UCNA) \cite{Maggi15}
and the Fox approximation method \cite{Brader}, as well as a
more controlled perturbative expansion \cite{Fodor16}.
We briefly describe below these different approaches.

\subsubsection{Modeling self-propulsion as a coloured noise}

The starting point is to model the self-propulsion force as a Gaussian coloured noise with a persistence time $\tau$. This is generically an approximation of standard self-propulsion forces, but it becomes exact in a limiting case where the direction of self-propulsion freely diffuses and the intensity of the force fluctuates with an appropriately chosen distribution. The persistence time $\tau$ is then proportional to the inverse of the angular diffusion coefficient of the force.

The theory deals with an arbitrary number $N$ of degrees of freedom
$(x_1,\dots,x_N)$. These may be interpreted as the components of the positions
vectors $({\bf r}_1,\dots,{\bf r}_{N'})$ of $N'=N/d$ particles
in a $d$-dimensional space, but the calculation presented is general and does not rely on this specific interpretation.

The overdamped dynamics satisfies the following equation of motion,
\be \label{eq:dyn:SPP:UCNA}
\frac{dx_i}{dt} = -\lambda \frac{\partial U}{\partial x_i} + \eta_i(t)
\ee
where $\lambda$ is the mobility coefficient, and $U(x_1,\dots,x_N)$ is a potential energy. The fluctuating term $\eta_i(t)$
is an exponentially correlated coloured noise satisfying
\be
\la \eta_i(t) \ra=0, \qquad \la \eta_i(t)\eta_j(t') \ra
= \delta_{ij} \frac{\lambda T}{\tau} e^{-|t-t'|/\tau} \,,
\ee
where $T$ is an effective temperature, introduced here to recover standard equilibrium results in the limit $\tau \to 0$.
In the following, we set $\lambda=1$ by an appropriate choice of units,
to lighten notations.

The exponentially correlated noise $\eta_i(t)$ can be interpreted as resulting from an Ornstein-Uhlenbeck process,
\be \label{eq:SPP:OU}
\frac{d\eta_i}{dt} = -\frac{\eta_i}{\tau} + \frac{1}{\tau} \xi_i(t)
\ee
where $\xi_i(t)$ is a white noise satisfying
\be
\la \xi_i(t) \ra=0, \qquad \la \xi_i(t)\xi_j(t') \ra
= 2T \delta_{ij} \delta(t-t')
\ee
(we recall that $\lambda=1$).

\subsubsection{Unified Coloured Noise Approximation}
\label{sec:UCNA}

We now look for a simple approximation which allows one to determine the
steady-state $N$-body distribution associated to the process defined in 
Eq.~(\ref{eq:dyn:SPP:UCNA}).
Taking the time derivative of Eq.~(\ref{eq:dyn:SPP:UCNA}) and combining the resulting equation with Eq.~(\ref{eq:SPP:OU}), one obtains
\be \label{eq:before:UCNA}
\tau \frac{d^2 x_i}{dt^2} + \sum_j \left( \delta_{ij}+ \tau \frac{\partial^2 U}{\partial x_i \partial x_j} \right) \frac{dx_j}{dt} = -\frac{\partial U}{\partial x_i} + \xi_i(t).
\ee
The UCNA consists in neglecting the second derivative term $\tau d^2 x_i/dt^2$
in Eq.~(\ref{eq:before:UCNA}) \cite{Hanggi87,Hanggi89,Luo93,Maggi15}.
The approximation, though strictly speaking uncontrolled, is assumed to be valid both for $\tau \to 0$ and
for $\tau \to \infty$ \cite{Maggi15,Maggi16}.
It is convenient to introduce the (position-dependent) symmetric matrix $M$ defined as
\be \label{eq:def:M:UCNA}
M_{ij} = \delta_{ij}+ \tau \frac{\partial^2 U}{\partial x_i \partial x_j} \,.
\ee
In the following, the matrix $M$ is assumed to be invertible;
the inverse matrix $M^{-1}$ is also symmetric.
Multiplying Eq.~(\ref{eq:before:UCNA}), in which the second order time derivative has been dropped, by the matrix $M^{-1}$ yields \cite{Maggi15}
\be \label{eq:Langevin:UCNA}
\frac{dx_i}{dt} = -\sum_j (M^{-1})_{ij}
\frac{\partial U}{\partial x_j} + \sum_j (M^{-1})_{ij} \xi_j(t)
\ee
which is a multiplicative Langevin equation, to be interpreted in the Stratonovich sense. This equation serves as the starting point in the derivation of
the joint distribution of positions, that we now address
(see also \cite{Szamel14} for the case of a single self-propelled particle in a potential).

The probability distribution $P(x_1,\dots,x_N,t)$ describing the Langevin equation (\ref{eq:Langevin:UCNA}) is governed by the following Fokker-Planck equation,
\be \label{eq:FP:UCNA}
\frac{\partial P}{\partial t} = -\sum_i \frac{\partial J_i}{\partial x_i}
\ee
where the probability current $J_i(x_1,\dots,x_N,t)$ is given by \cite{Maggi15}
\be \label{eq:Ji:UCNA}
J_i = -\sum_j  (M^{-1})_{ij} \left( \frac{\partial U}{\partial x_j} P
+ T \sum_{k} \frac{\partial}{\partial x_k} [(M^{-1})_{jk} P] \right) .
\ee
Since we are not considering systems with alignment interactions, one does not expect to observe a macroscopic particle current in steady state.
However, there could be stationary probability currents in phase space, since we are dealing with a non-equilibrium system.
Yet, following \cite{Maggi15}, it is possible to consider the stationary solution of Eq.~(\ref{eq:FP:UCNA}) with vanishing current, $J_i=0$.
After some relatively straightforward algebraic manipulations, and using
Jacobi's formula
\be
\frac{\partial}{\partial x_i} \ln |{\rm det} M|
= {\rm Tr}\left( M^{-1} \frac{\partial M}{\partial x_i} \right),
\ee
the condition $J_i=0$ can be reformulated as
\be
\frac{\partial P}{\partial x_i} =
\left( -\frac{1}{T} \sum_j M_{ij} \frac{\partial U}{\partial x_j}
+ \frac{\partial}{\partial x_i} \ln |{\rm det} M| \right) P \,.
\ee
Recalling the definition (\ref{eq:def:M:UCNA}) of the matrix $M$,
the stationary distribution $P(x_1,\dots,x_N)$ is obtained as
\be \label{eq:dist:xi:UCNA}
P = \frac{1}{Z} \, |{\rm det} M|
\exp\left[ -\frac{U}{T} - \frac{\tau}{2T}
\sum_i \left( \frac{\partial U}{\partial x_i} \right)^2 \right]
\ee
with $Z$ a normalization constant.
Note that the explicit dependence of $P$, $U$ and $M$
on the variables $(x_1,\dots,x_N)$ has again been dropped to lighten notations.
Eq.~(\ref{eq:dist:xi:UCNA}) generalizes the Boltzmann-Gibbs probability distribution, and reduces to it in the limiting case $\tau=0$.

It is possible to reformulate Eq.~(\ref{eq:dist:xi:UCNA}) using vectorial notations, which are both more compact and more explicit.
Defining the $N$-dimensional vector ${\bf x}=(x_1,\dots,x_N)$
and the $N$-dimensional gradient
\be
{\boldsymbol \nabla}_{\!\! N} \equiv \left( \frac{\partial}{\partial x_1},\dots,
\frac{\partial}{\partial x_N} \right),
\ee
Eq.~(\ref{eq:dist:xi:UCNA}) can be rewritten as \cite{Maggi15}
\be \label{eq:Px:UCNA}
P({\bf x}) = \frac{1}{Z} \left| {\rm det}\big( {\bf I}+\tau {\boldsymbol \nabla}_{\!\! N} {\boldsymbol \nabla}_{\!\! N} U({\bf x}) \big) \right| \,
\exp \left[ -\frac{U({\bf x})}{T} - \frac{\tau}{2T}
\big( {\boldsymbol \nabla}_{\!\! N} U({\bf x}) \big)^2 \right] .
\ee
In the small $\tau$ limit, Eq.~(\ref{eq:dist:xi:UCNA}) can be expanded to first order in $\tau$, yielding an effective equilibrium distribution \cite{Fodor16}
\be \label{eq:eff-equil:UCNA}
P({\bf x}) = \frac{1}{Z} \, e^{-U_{\rm eff}({\bf x})/T}
\ee
with an effective potential energy
\be \label{eq:SPP:Ueff}
U_{\rm eff}({\bf x}) = U({\bf x}) + \tau \sum_i \left[ \frac{1}{2} \left( \frac{\partial U}{\partial x_i} \right)^2
- T \frac{\partial^2 U}{\partial x_i^2} \right] +\mathcal{O}(\tau^2) \,.
\ee
The effective potential (\ref{eq:SPP:Ueff}) can then serve as a basis to use standard tools of liquid theory like integral equations, which allows for a direct comparison with numerical simulations \cite{Brader,Maggi15SM}.
For system with a purely repulsive potentiel $U$, the theory predicts
a phase separation induced by the persistence time $\tau$ \cite{Brader}, thus providing a microscopic (though approximate) theoretical foundation for Motility Induced Phase Separation \cite{Tailleur-rev}.
The advantage of starting from the (non-factorized) $N$-body distribution
is precisely to be able to compute correlations between particles, which cannot be accessed through standard local mean-field or Boltzmann equation approach,
since these methods focus on the one-body distribution\footnote{However, note that more general kinetic theory approaches can capture two-body (or even higher order) correlations,
by using higher order truncations of the BBGKY hierarchy (see, e.g., \cite{Ihle15}).}.
Pair correlations predicted from the effective potential approach have been found to show quantitative agreement with numerical simulations \cite{Brader}.

The effective equilibrium distribution given in Eq.~(\ref{eq:eff-equil:UCNA}) suggests that the system can be effectively described as an assembly of Brownian particles with an effective pair potential which depends on the correlation time $\tau$.
However, the practical range of validity of this approximation partly remains an open issue. Very recent numerical simulations indicate
that the effective potential approximation (\ref{eq:SPP:Ueff})
may be valid only in a very limited range of small values of the persistence time $\tau$ and of the self-propulsion speed $v_0$ \cite{Speck16}.
Moreover, these numerical results also indicate that the mapping to
Brownian particles with an effective pair potential fails to describe dynamical quantities like the Virial pressure, even for small $\tau$ and $v_0$ \cite{Speck16}.
Further work is certainly needed to clarify, at the theoretical level, the precise range of validity of these approximation schemes, and how they could possibly be improved to describe dynamical quantities.

\subsubsection{Fox approximation for Langevin equations with coloured noise}

An alternative method, called the Fox method, has also been proposed to obtain an approximate expression of the distribution $P({\bf x})$ for Langevin equations with exponentially coloured noise \cite{Brader,Fox1,Fox2}.
As we have seen in Sect.~\ref{sec:UCNA}, the UCNA approximates the dynamics and determines the $N$-body distribution $P({\bf x})$ corresponding to the approximate dynamics.
Instead, the Fox method consists in writing an exact formal equation for the time evolution of the $N$-body distribution $P({\bf x})$ corresponding to the `true' dynamics. This exact equation is not closed in terms of $P({\bf x})$,
but it can be turned into a closed Fokker-Planck equation using a simple approximation \cite{Brader,Fox1}.
As the resulting stationary distribution turns out to be the same as the one obtained from UCNA \cite{Fodor-thesis},
we will not provide a detailed derivation of this stationary distribution in the $N$-body case, but simply discuss the Fox method in the case of a single degree of freedom subjected to a conservative force and a coloured noise \cite{Brader,Fox1}.
We thus consider the following dynamics,
\be \label{eq:Langevin:Fox}
\frac{dx}{dt} = -U'(x)+\eta(t)
\ee
where $U(x)$ is a single-particle potential energy, and the prime denotes the derivative. The noise $\eta$ is an exponentially correlated Gaussian noise satisfying
\be
\langle \eta(t) \rangle =0, \qquad
\langle \eta(t)\eta(t') \rangle = C(t-t') \equiv \frac{T}{\tau}
\, e^{-|t-t'|/\tau} \,.
\ee
The distribution of a noise `trajectory'
$\eta(t)$ over a given time interval $t_1<t<t_2$ can be written as a Gaussian functional
\be \label{eq:Peta:Fox}
\mathcal{P}[\eta] \propto \exp\left( -\frac{1}{2} \int_{t_1}^{t_2} dt
\int_{t_1}^{t_2} dt' \, K(t-t') \eta(t) \eta(t') \right)
\ee
where the kernel $K(t-t')$ is the inverse of the correlation $C(t-t')$
in the convolution sense:
\be
\int_{-\infty}^{\infty} ds \, K(t-s)\, C(s-t') = \delta(t-t') \,.
\ee
The distribution $P(x,t)$ can be formally written as
\be
P(\tilde{x},t) = \langle \delta\big(\tilde{x}-x(t) \big) \rangle_{[\eta]}
\ee
where $\langle \cdots \rangle_{[\eta]}$ denotes an average over noise trajectories with a probability given by Eq.~(\ref{eq:Peta:Fox}).
Note that we have used the notation $\tilde{x}$ to distinguish it from the
time-dependent solution $x(t)$ of Eq.~(\ref{eq:Langevin:Fox}).
Using functional calculus as well as the Gaussian property of the noise $\eta$,
one can derive a formal equation for $P(\tilde{x},t)$ \cite{Brader,Fox1},
\bea \nonumber
\frac{\partial P}{\partial t}(\tilde{x},t) &=& 
\frac{\partial}{\partial \tilde{x}} \big( U'(\tilde{x}) P(\tilde{x},t) \big)\\
&+& \frac{\partial^2}{\partial \tilde{x}^2} \left[
\int_0^t dt' \, C(t') \left< e^{-\int_{t-t'}^t ds \, U''(x(s))}
\delta\big(\tilde{x}-x(t) \big) \right>_{[\eta]} \right] .
\label{eq:Fox:formal}
\eea
Although this equation bears some resemblence with a Fokker-Planck equation,
it is not a closed equation in terms of the probability $P(x,t)$,
and it is thus, as it stands, of little use for practical purposes.
Hence, the idea is to perform an approximation on Eq.~(\ref{eq:Fox:formal})
in order to write it in a closed form.
This can be easily done as follows.
Assuming that $x(s)$ does not vary significantly over the correlation time of the noise, the integral appearing in the exponential in Eq.~(\ref{eq:Fox:formal}) can be approximated, to lowest order in $t'$, as
\be
\int_{t-t'}^t ds \, U''\big(x(s)\big) \approx U''\big(x(t)\big) \, t' \,.
\ee
With this approximation, Eq.~(\ref{eq:Fox:formal}) can be rewritten as a Fokker-Planck equation, which for $t \gg \tau$ simply reads (dropping the tildes to lighten notations)
\be
\frac{\partial P}{\partial t}(x,t) =
\frac{\partial}{\partial x} \left[
 U'(x) P(x,t) + T \frac{\partial}{\partial x}
\left( \frac{P(x,t)}{1+\tau U''(x)} \right) \right],
\ee
where we have used the explicit exponential form of the noise correlation.
The probability current $J(x)$, defined by
$\partial P/\partial t = -\partial J/\partial x$ then reads
\be \label{eq:Fox:current}
J(x) = - U'(x) P(x,t) - T \frac{\partial}{\partial x}
\left( \frac{P(x,t)}{1+\tau U''(x)} \right).
\ee
Eq.~(\ref{eq:Fox:current}) is to be compared with the probability current obtained from UCNA for a single degree of freedom [see Eq.~(\ref{eq:Ji:UCNA})],
\be
J_{\rm UCNA}(x) = -\frac{1}{1+\tau U''(x)} \left[ U'(x) P(x,t) + T \frac{\partial}{\partial x}
\left( \frac{P(x,t)}{1+\tau U''(x)} \right) \right].
\ee
One thus has $J_{\rm UCNA}(x)=J(x)/[1+\tau U''(x)]$, so that the Fox and UCNA methods lead to the same steady state distribution, as determined by the zero flux condition. The dynamics resulting from both approximations is however different,
showing that the methods are not equivalent.

For the full $N$-body problem, the current $J_i$ 
obtained from the Fox approximation
reads\footnote{The result originally published in \cite{Brader} has a slightly different expression, but it was argued by other authors that the derivation given in \cite{Brader} contained an error \cite{Fodor-thesis,Cates-journal-club}. We report here the corrected expression given in \cite{Fodor-thesis}.} \cite{Fodor-thesis}
\be \label{eq:current:UCNA:Nbody}
J_i = -  \frac{\partial U}{\partial x_i} P
- T \sum_{k} \frac{\partial}{\partial x_k} [(M^{-1})_{ik} P]
\ee
where the position-dependent matrix $M_{ij}$ has been defined in Eq.~(\ref{eq:def:M:UCNA}).
The current (\ref{eq:current:UCNA:Nbody}) differs from the current obtained from UCNA, as given by Eq.~(\ref{eq:Ji:UCNA}), only by a global matrix prefactor.
Hence the zero flux condition also leads to the same steady-state distribution for both the Unified Coloured Noise and Fox approximations in the $N$-body case.
Comparison of the UCNA and Fox method results to numerical simulations in a time-dependent regime would thus be useful to assess the validity of these two approximation schemes.

\subsubsection{Perturbative expansion for small persistence time}

The main drawback of both the UCNA and the Fox method is that they are essentially uncontrolled.
The UCNA has been assumed to be valid both in the limits of small and large values of the persistence time $\tau$ \cite{Hanggi89,Maggi15}
---although the approximation made to obtain Eq.~(\ref{eq:Langevin:UCNA})
is more intuitively understood as a small $\tau$ approximation.
The Fox method assumes that the position of the particles does not
significantly vary over a duration $\tau$, suggesting here also that 
$\tau$ should be small.
Hence, it is natural to try to determine the $N$-body
distribution using a systematic small-$\tau$ expansion \cite{Fodor16}.
This expansion requires to consider the joint distribution 
$\tilde{P}({\bf x},{\bf v})$ of positions and velocities
(we recall that ${\bf x}$ and ${\bf v}$ are here $N$-dimensional vectors gathering all degrees of freedom in the system).
In terms of the variables ${\bf x}$ and ${\bf v}$,
Eq.~(\ref{eq:before:UCNA}) can be reformulated as
coupled first order stochastic differential equations,
\bea \nonumber
&\frac{dx_i}{dt}& = v_i \\
\tau &\frac{d v_i}{dt}& = -\sum_j \left( \delta_{ij}+ \tau \frac{\partial^2 U}{\partial x_i \partial x_j} \right) v_j -\frac{\partial U}{\partial x_i} + \xi_i(t)
\eea
for which a Fokker-Planck equation can be written.
A systematic expansion of this Fokker-Planck equation in terms of the small parameter $\sqrt{\tau}$ can then be performed \cite{Fodor16}.
Note that the expansion requires a rescaling of velocities, by defining $\tilde{v}_i = \sqrt{\tau} v_i$.
Although the expansion parameter is $\sqrt{\tau}$ instead of $\tau$, the parameter $\sqrt{\tau}$ does not explicitly appear at leading order, because
the expansion of the probability distribution $P({\bf x})$ starts at second order, that is at order $\tau$.
One then finds at order $\tau$ for the joint distribution
of positions and velocities \cite{Fodor16}
\be \label{eq:Nbody:Prv}
\tilde{P}({\bf x},{\bf v}) = \frac{1}{Z} \, \exp \left\{
\frac{1}{2T} \sum_i \tilde{v}_i^2 + \frac{U}{T}
-\frac{\tau}{2} \sum_i \left[ \left( \frac{\partial U}{\partial x_i} \right)^2
+ \left( \tilde{v}_i \frac{\partial}{\partial x_i}\right)^2 U
-3T \frac{\partial^2 U}{\partial x_i^2} \right] \right\}
\ee
(note that we generically use the same notation $Z$ for all normalization constants). Higher order corrections, starting with terms of order $\tau^{3/2}$,
can also be systematically derived \cite{Fodor16}.
Using the rescaled velocity variables $\tilde{v}_i$ allows for a clearer presentation of the result in terms of an expansion in $\sqrt{\tau}$.
Note that restoring the original velocities $v_i$, one finds in particular that the mass of the particles, appearing in the standard kinetic energy term, is equal to $\tau$ in the units used.

In order to compare with results predicted from UCNA, one may compute the conditional $N$-body distribution of velocities at a fixed position $P({\bf v}|{\bf x})$.
At order $\tau$ in the perturbative expansion, one obtains a Gaussian distribution \cite{Fodor16}, more conveniently written using the $N$-dimensional vector notations,
\be \label{dist:pvx:UCNA}
P({\bf v}|{\bf x}) = \frac{1}{Z}
\exp\left( -\frac{\tau}{2T}\, {\bf v} \cdot [ {\bf I}+ \tau
{\boldsymbol \nabla}_{\!\! N} {\boldsymbol \nabla}_{\!\! N} U] \cdot {\bf v} \right)
\ee
(we have used here the physical velocities ${\bf v}$, and not the scaled velocities ${\bf \tilde{v}}$).
Eq.~(\ref{dist:pvx:UCNA}) is identical to the result obtained in the framework of UCNA \cite{Maggi16}.
Taking into account higher order corrections, starting with terms of order $\tau^{3/2}$, the distribution (\ref{dist:pvx:UCNA}) becomes non-Gaussian \cite{Fodor16}.

A further comparison with UCNA results is obtained by computing the $N$-body distribution of positions $P({\bf x})$, which can be deduced from
$\tilde{P}({\bf x},{\bf v})$ by integrating over the velocities.
To order $\tau$, one recovers the UCNA result given in Eq.~(\ref{eq:Px:UCNA}) \cite{Fodor16}. Differences are however expected to appear at higher order in $\tau$.


\subsection{Discussion}

General methods to determine the $N$-body stationary distribution
of non-equilibrium systems are scarce.
We have reviewed some of them in this section, being aware that each of them has a relatively limited range of applicability, although it addresses a class of system and not a single model.
In practice, exact solutions for stochastic models without detailed balance can be found only for some specific cases.
Non-factorized solutions are known mostly for one-dimensional models, often
through Matrix Product Ansatz solutions \cite{Evans-MPArev}
(see Sect.~\ref{sec-exact-solv}).
Note that models with pair-factorized solutions have also been proposed
\cite{Evans06}.
Even for one-dimensional systems with local stochastic rules, 
for which the existence of a Matrix Product solution is granted
\cite{Sandow97}, it is often hard to determine the solution explicitly.
More general formal solutions exist, in particular in the framework of the
McLennan ensembles \cite{McLennan,Komatsu08,Komatsu09,Maes10}, but here again, the applicability of the method to practical situations has been up to now very limited.

Approximate solutions like the Edwards approach for dense granular matter
(Sect.~\ref{sec:Edwards-Nbody}) or
the Unified Coloured Noise approximation (Sect.~\ref{sec:SPP-Nbody})
a priori have a broader range of applicability than exact solutions, since they do not strongly depend on details of the model.
However, their practical use may require further approximations to deal with complicated interactions, as the constraints imposed by mechanical stability for instance can hardly be treated exactly in dimension higher than one
(see Sect.~\ref{sec-levine-grains}).

Interestingly, exactly solvable models may in some cases shed some light on the type of approximation that could be relevant to some classes of systems, and on their range of validity.
For instance, the exact solution of the boundary driven Zero Range Process
presented in Sect.~\ref{sec-ZRP} can be generalized by including a site-dependent rate of particle transfer as well as a tree geometry, with the aim to model an energy cascade in a driven-dissipative system. In this picture, particles are interpreted as energy amounts,
sites correspond to different length scales, and reservoirs to energy injection at large scale and dissipation at small scale.
The exact solution shows that depending on the energy transfer mechanism, the stationary distribution may converge, in the limit of a large separation of scales between injection and dissipation, either to a quasi-equilibrium distribution or to a fully dissipative state with a finite dissipated flux \cite{DauchotPRL09}. Such exact solutions may thus give hints of when an effective equilibrium distribution may be relevant in a dissipative system: systems with low dissipated flux are more likely to be described by an equilibrium-like distribution.
Although this result does not come as a surprise, the most interesting point resulting from the exact solution of the Zero Range Process is that there is a sharp transition between quasi-equilibrium and fully dissipative regimes
as a function of a parameter characterizing the internal energy transfer
\cite{DauchotPRL09}.
This suggests that there might be classes of systems with each behaviour.
This is to some extent reminiscent of the behaviour of fluid turbulence, which although much more complex than the Zero Range Process, is weakly dissipative in two dimensions, and strongly dissipative in three dimensions \cite{Frisch}.
Equilibrium-like approaches are thus more suitable for the two-dimensional case \cite{Bouchet-review}.


\section{Conclusion and outlook}
\label{sec-conclusion}

In this review, we have tried to present a brief overview of the statistical methods that are available to describe the statistics of large assemblies of interacting dissipative units. Examples of such units include inelastic granular particles, self-propelled particles, bubbles in foams, plastic events in elastoplastic systems, low-dimensional dynamical systems, or even more abstract objects like Fourier or normal modes.
We have devoted a large part of the review (Sect.~\ref{sec-MF} to \ref{sec-kinetic-th}) to methods that reduce, at the cost of different types of approximations, the complexity of the problem to the study of a single unit interacting with a self-consistent environment.
Such generic types of methods include elementary mean-field theory (either static or dynamic) as well as more involved local descriptions which, although focusing on the dynamics of a single unit, manage to capture relevant spatial information, as in the local mean-field approach and in kinetic theory.
It is also worth noticing that kinetic theory, in the form of the Boltzmann equation, is supposed to become exact in the low density limit (except when noise is simultaneously decreased with density, as in the case of self-propelled particles with velocity alignment \cite{Hanke13}), which distinguishes it from mean-field approaches that mostly rely on uncontrolled approximations.

To go beyond the description of a single unit, we have then focused in Sect.~\ref{sec-Nbody} on approaches based on the determination of the full $N$-body distribution.
For some specific classes of one-dimensional models, exact solutions for the 
$N$-body distribution can be found, either in the simple form of a factorized distribution as in the Zero Range Process (and in similar types of mass transport models), or in the more complicated form of matrix products as in the ASEP and in some reaction-diffusion lattice models.
However, the class of models that can be dealt with using such methods is limited; In particular, the Matrix Product Ansatz is restricted by construction to one-dimensional lattice models.
For other types of models, one has to resort to approximation schemes,
and we have discussed some of them in Sect.~\ref{sec-Nbody}, trying to emphasize some general enough approaches.
We have discussed in particular the Edwards approach for dense granular matter (thereby going beyond the simple mean-field treatment presented in Sect.~\ref{sec-MF}), and approximation schemes that are relevant for Langevin equations with coloured noise, with application to systems of interacting self-propelled particles.

All along this review, we have tried to follow a clear line of thought, going from the single unit approximation to the full system treatment in a consistent
progression of approaches, illustrated on several explicit examples to show the generality of each type of method. Accordingly, we have put a strong emphasis on methodological aspects related to the statistical description of systems made of a large number of interacting dissipative units, leaving aside most of the specificities of each system considered.

We have also tried to present some physical problems using different complementary approaches, to see the advantages and limitations of each of them, and which kind of information each approach may provide.
This was the case for dense granular matter, for foams and for self-propelled particles with velocity alignment interactions.
For dense granular matter (or more generally driven athermal systems with dry friction), we have seen within the framework of the Edwards hypothesis how a very simple mean-field approach, neglecting correlations, already yields a description of the segregation of grains with different friction coefficients (Sect.~\ref{sec-levine-grains}).
Still using the Edwards approach,
a more sophisticated transfer operator treatment of a one-dimensional spring-block model leads to the description of growing correlations when increasing the driving (Sect.~\ref{sec:Edwards-Nbody}).
For complex fluids, we have seen, focusing on stress statistics, how a phenomenological mean-field treatment (Sect.~\ref{sec-HL}) can be improved using the local mean-field approximation (Sect.~\ref{sec-KEP}), yielding predictions for the phenomenological parameters
and allowing for a space-dependent description, which may be important to account for boundary effects for instance.
Finally, self-propelled particles with velocity alignment interactions
have been described in the framework of both the local mean-field approximation (Sect.~\ref{sec-SPP-FP}) and the Boltzmann equation (Sect.~\ref{sec-SPP-Boltz}). The resulting continuous equations for the velocity field have the same form, being constrained by symmetries.
The functional dependence of coefficients on the density is however different, leading to quantitative differences in the phase diagrams, that can be compared with particle-based simulations. The Boltzmann equation is expected to be more accurate at low density, while the local mean-field approximation should rather describe situations where the interaction range is large as compared to the particle size. Besides, in the absence of alignment interactions,
self-propelled particles can also be described by their $N$-body probability distribution, in the framework of coloured noise approximations (Sect.~\ref{sec:SPP-Nbody}). Such an approach yields at lowest order in the persistence time an effective pair potential, allowing for the use of methods from equilibrium liquid theory to describe the system.

Since this review was aimed at being relatively concise,
some important topics have not been covered.
These include, for instance, the statistical physics of turbulent flows \cite{Frisch}, a topic of broad interest in which significant progress has been made over the last decades \cite{Bouchet-review}.
A more recent development that has not been addressed either is the generalization of the Mode Coupling Theory (a theory aiming to study the glass transition starting from a microscopic description, see e.g.~\cite{Charbonneau}) to dense assemblies of self-propelled particles \cite{Berthier15,Szamel16}.
Such an approach is certainly promising in order to understand how self-propulsion \cite{Henkes11,Henkes14,Berthier14a,Berthier14b,Berthier16a}, or other types of activity \cite{Berthier16b}, may modify the glassy properties of a dense system.

Others classes of problems that are to some extent related to the scope of this review are the problems of growing interfaces as described by the Edwards-Wilkinson \cite{EW82} and Kardar-Parisi-Zhang (KPZ) \cite{KPZ86,Corwin12} equations, and that of interfaces driven through a disordered medium \cite{Rosso01,Giamarchi06,Ferrero13}.
One of the most efficient ways to tackle such interface growth  problem is the renormalization group approach \cite{Wiese98,Canet10},
a method (initially developed for equilibrium critical phenomena \cite{textbook-RG}) in which small scale degrees of freedom are progressively integrated out to determine the large scale behaviour.
The renormalization group method
has also been applied to driven manifolds in random media \cite{Nattermann92,Fisher93,LeDoussal02}, as well as to reaction-diffusion processes \cite{Lee94,Cardy95a,Cardy95b,Hinrichsen00}.
In addition, it has recently been applied to the study of fully developed turbulence in the Navier-Stokes equation, yielding very promising results \cite{Canet16a,Canet16b}.

Among the many perspectives of the still very open field of statistical physics of dissipative units, one might guess that future developments may be more centered on the formalism of large deviation functions \cite{Touchette-rev,Ellis-book}.
This mathematical tool plays a unifying role in statistical physics,
in the sense that it is the relevant probabilistic framework to formulate the statistics of systems having many degrees of freedom, whether at or out of equilibrium \cite{Ellis99}.
Large deviation functions have been used for instance to generalize to out-of-equilibrium systems the notion of free energy, both in the context of boundary driven systems \cite{Derrida01,Derrida02} and of active systems \cite{Barre}.
Along this line, the development of a generic formalism based on large deviation approaches would be desirable for dissipative systems.


\bigskip


\end{document}